\documentclass[aoas]{imsart}
\RequirePackage{amsthm,amsmath,amsfonts,amssymb}
\RequirePackage[authoryear]{natbib}

\usepackage{lipsum}
\usepackage{tabu}
\usepackage{adjustbox}
\usepackage[english]{babel}
\usepackage[utf8]{inputenc}
\usepackage{longtable}
\usepackage{amsmath}
\usepackage{hyperref}
\usepackage{graphicx}
\usepackage{enumerate}
\usepackage{enumitem}
\usepackage[colorinlistoftodos]{todonotes}
\usepackage[algo2e]{algorithm2e}
\usepackage{algorithmic}  
\usepackage{algorithm}
\usepackage{comment}
\usepackage{array}
\usepackage{natbib}

\begin{document}

\begin{frontmatter}
\title{A Dynamic Additive and Multiplicative Effects Network Model
	with Application to the United Nations Voting Behaviors \thanksref{t1}}
\runtitle{}
\thankstext{T1}{This work was supported by the National Institute of Allergy and Infectious Diseases of the National Institutes of Health under award number R01AI136664.}
\begin{aug}
\author[A]{\fnms{Bomin} \snm{Kim}},
\author[B]{\fnms{Xiaoyue} \snm{Niu}},
\author[B]{\fnms{David} \snm{Hunter}}
\and
\author[B]{\fnms{Xun} \snm{Cao}}
\address[A]{Freddie Mac} 
\address[B]{The Pennsylvania State University}
\end{aug}
 
\begin{abstract}
	
Motivated by a study of United Nations voting behaviors, we 
introduce a regression model for a series of networks that are 
correlated over time. Our model is a dynamic extension of the additive 
and multiplicative effects network model (AMEN) of \cite{hoff2021additive}. In 
addition to incorporating a temporal structure, the model accommodates 
two types of missing data thus allows the size of the network to vary 
over time. We demonstrate via simulations the necessity of various 
components of the model. We apply the model to the United Nations 
General Assembly voting data from 1983 to 2014 \citep{voeten13} to 
answer interesting research questions regarding international voting 
behaviors. In addition to finding important factors that could explain 
the voting behaviors, the model-estimated additive effects, 
multiplicative effects, and their movements reveal meaningful foreign 
policy positions and alliances of various countries. 
\end{abstract}


\end{frontmatter}

\section{Introduction} \label{sec:Introduction}

The United Nations was founded in 1945 and
today it includes as members nearly all of the world's sovereign states.  Each
UN member state is also a member of the
United Nations General Assembly (UNGA), the main deliberative body of the UN,
and each is eligible to vote on matters that come before the UNGA.  The votes of
the member states reveal much about various states' stances 
towards each other and on various issues, which helps to understand past 
foreign policies and state behaviors and predict those yet to come. 
As such, these votes have been analyzed in 
many political science papers \citep[see, e.g.,][and the references therein]{voeten13} 
and have become standard data sources to study states' preferences, 
one of the most important topics in the field of international relations.

This article combines and extends statistical ideas to create
a novel methodology, then applies this methodology to
the UNGA voting data in three innovative ways.
First, we consider voting behavior to be dyadic, that is, we explicitly model the 
similarity and differences in nations' pairs of votes by expressing the voting data in the form
of a network dataset that captures the degree to which pairs of nations agree in their votes.
Other researchers have viewed UN voting as 
dyadic behavior and have thus used dyadic similarity indicators such as 
affinity or S scores \citep[e.g.,][]{gartzke1998kant,signorino1999tau}, but to our 
knowledge, the United Nations voting data have never been analyzed using 
modern statistical network models.
Second, we explicitly model the time dependence of voting agreement.
Votes are of course highly correlated over time, because they are the reflections of history. 
Our model allows for more general temporal dependence than the methods that
have previously attempted to account for this dependence using a Markovian
assumption \citep[e.g., the dynamic ordinal spatial item response theory model of][]{bailey2017estimating} 
whereby temporal dependence lasts a single time step, i.e., one year.
Third, by considering voting patterns as a network dataset in which each nation state 
is assumed to occupy a position in latent space that may evolve over time, 
our methodology allows the analysis of high-order dependences such as transitivity and 
stochastic equivalence.
This allows us to consider the fact that 
voting decisions are not limited to dyadic relations---country A's 
decision to vote along with country B might well be influenced by 
country C's decision.

We refer to our modeling framework as the dynamic additive 
and multiplicative effects (DAME) model, since we take as its basis the additive and multiplicative
effects (AME) framework of \cite{hoff2021additive}.  We extend the AME framework
to handle time-varying networks
using Gaussian process dependence structure to arrive at DAME.  Our model allows for different
forms of the Gaussian process covariance structure, whose length scale we estimate from the data,
as well as various types of missing data.
In Section \ref{sec:DAME}, we introduce the DAME model, derive the 
sampling distribution for Bayesian inference, and describe the treatment 
of missing data. Then we present simulation studies to show the necessity 
of various components in the DAME model. And in Section \ref{sec:UNvoting} 
we apply the DAME model to the United Nations General Assembly 
voting network, demonstrate the validity of the model, and discuss the 
findings.

\section{Dynamic Additive and Multiplicative Effects Model}
\label{sec:DAME}

Networks can be used to describe relationships among subjects---in this 
case, members of the UN---by representing
the subjects as the nodes of a network and the relationships as 
the edges of the network. 
In some contexts, the relationship might be binary, 
as the presence of political alliances between country A and country B,
or directed, as the indicator that country A initiated a conflict with
county B.  In the current context, we consider a particular relationship 
that is quantitative and undirected, namely, the voting agreement between
pairs of countries.

There has been a lot of work in developing statistical models for networks.
With this development, along with the increasing availability of network data,
there has been a growing need and capability to model networks
that change over time.  Some models are intrinsically dynamic in nature,
such as the stochastic actor-oriented models originated by \cite{snijders01}
and developed in subsequent articles \citep[e.g.,][]{snijders2010introduction}
that envision ties as random processes that evolve in time.
Other work involves extensions of static network models into the time-varying
domain.  For instance, the sizable literature on 
exponential-family random graph models (ERGMs) 
\citep[see, e.g.,][]{lusher2013} has been extended by papers such as
\cite{hanneke2010discrete} and \cite{stergm} to handle discrete-time
time-varying networks in which a network at time $t+1$ depends explicitly 
on the network at time $t$ via one or more statistics measured on the 
network.

One of the features of social networks and in general human behaviors is that we cannot observe all factors that determine the behaviors. Latent variables help capture this feature and are very common in social science models. An important subset of the statistical networks literature, to which the current article belongs, 
involves latent variables, i.e., unobserved properties of the network. 
\cite{kim2017review} provide a comprehensive review of the literature on dynamic 
network models with latent variables, and more recent development includes \cite{dsbm22}. Here, we consider one type of the latent variable models, i.e. the latent space models,
which include distance and projection models 
\citep{hoff2002latent}, the bilinear effects model 
\citep{hoff2005bilinear}, and the additive and multiplicative effects 
model (AME) \citep{hoff2021additive}. 
While there exist dynamic extensions of the latent 
distance model and bilinear effects model, to the
best of our knowledge, the AME model has not been extended to the dynamic 
case, which is our focus in this article. 


Another challenge in dynamic network modeling is missing data, including 
missing edges and missing nodes over time. Common practices for handling 
missing data are to use only the subset of nodes that have complete 
observations or to treat all missing data as missing at random and then 
imputing values to complete the dataset. It is not uncommon for some
countries' voting patterns to be disrupted by specific geopolitical 
factors that may persist over multiple years, so missing data should be 
treated carefully so as not to inadvertently introduce biases by assuming 
that such missingness occurs haphazardly. We argue in 
Section~\ref{subsec:VaryingNumberOfNodes}
that some of the missing mechanisms might not be 
missing at random, particularly those involving nodes missing for a stretch of time. 
We therefore propose novel treatments for different missing mechanisms.   

\subsection{Extending the AME framework}
\label{sec:ExtendingAME}

This article extends the Gaussian additive and multiplicative effects 
(AME) framework of \cite{hoff2021additive}.  Since the dataset we study in this
paper involves a symmetric network, our starting point is the symmetric AME, which assumes
\begin{equation}\label{AMEsymm}
\begin{aligned}
	&Y_{ij} = \sum_{p=1}^P{X}_{ijp}{\beta}_p+a_i + a_j + 
	\boldsymbol{u}^\prime_i \mathbf{D} \boldsymbol{u}_j + \epsilon_{ij},
\end{aligned}
\end{equation}
where $Y_{ij}$ is the $(ij)$th entry of the sociomatrix 
($\mathbf{Y}_{n \times n}$) of the network; $\boldsymbol{\beta}=(\beta_1, ..., 
\beta_P)$ is the $P$--length coefficient vector of predictor 
($\mathbf{X}_{n \times n \times P}$) effects (including the intercept); $a_i$ 
and $a_j$ are the nodal additive effects, which can be interpreted as each
node's ``sociability" or ``popularity"; $\boldsymbol{u}_i$ and $\boldsymbol{u}_j$ are
the $R$-dimensional latent factors for nodes $i$ and $j$; and
$\mathbf{D}$ is an $R\times R$ 
diagonal matrix, and finally $\epsilon_{ij}$ are random errors that are not captured by the previous terms. 
The asymmetric version of AME can be extended in a similar fashion, as we
discuss in Section~\ref{sec:Discussion}. 
The multiplicative 
effect $\boldsymbol{u}^\prime_i \mathbf{D} \boldsymbol{u}_j$
represents the similarity between node $i$ and node $j$, and in this expression
the signs of the entries of 
$\mathbf{D}$ have important substantive implications:
Positive and negative entries
indicate homophily and anti-homophily, respectively, with regard to the 
corresponding dimension in the latent $\boldsymbol{u}$ vectors.  That is, 
when nodes $i$ and $j$ have values of $u_{ir}$ and $u_{jr}$ that are large in 
magnitude and have matching signs, then the value of the corresponding diagonal
element $d_r$ of $\mathbf{D}$ determines whether this $r$th dimension contributes a 
positive or negative value to the mean of $Y_{ij}$.  A positive value corresponds to 
homophily, whereas a negative value corresponds to anti-homophily.


In extending the symmetric AME model to the time-varying domain,
we adapt an idea of \cite{durante2014} by 
assuming a Gaussian process  
on the time-varying components.   This approach allows for long-term dependence, unlike 
any model employing a Markovian assumption, since we know that countries'
behaviors towards one another fall into patterns that do not change 
dramatically from year to year. \cite{durante2014} extends a simplified version of Model~(\ref{AMEsymm}) in which $\boldsymbol{u}^\prime_i \boldsymbol{u}_j$ is used
in place of $\boldsymbol{u}^\prime_i \mathbf{D}  \boldsymbol{u}_j$. As discussed above, the $\mathbf{D}$ matrix allows flexibility in modeling both homophily and anti-homophily, and we further demonstrate in Section \ref{sec:simulation study} the necessity of including $\mathbf{D}$ in symmetric relational data.

Our dynamic additive and multiplicative effects (DAME) model assumes that
for $1\le j < i\le N$, $1\le t\le T$, $1\le p\le P$, and $1\le r\le R$,
\begin{eqnarray}\label{eqn:dame}
	Y^t_{ij} &=&\sum_{p=1}^P 
			{X^t_{ijp}}\beta^t_{p}+a^t_{i}+a^t_{j}+{{\boldsymbol{u}^t_{i}}^\prime 
			\mathbf{D}^{t} \boldsymbol{u}^t_{j}}+\epsilon^t_{ij},\\
	\boldsymbol{\beta}_{p}= (\beta^1_p,\ldots, \beta^T_p) &\sim& \mathcal{N}_T(\boldsymbol{0}, 
					\Sigma^\beta_p) \mbox{ with }
					\Sigma^\beta_p (t,t') = \tau_p^\beta f(\kappa^\beta_{p},t,t'), \nonumber\\
	\boldsymbol{a}_{i}=(a^1_i,\ldots, a^T_i)&\sim& \mathcal{N}_T(\boldsymbol{0},\Sigma^a) \mbox{ with }
					\Sigma^a(t,t') = \tau^a f(\kappa^a,t,t'), \nonumber\\
	\boldsymbol{u}_{ir}= (u^1_{ir},\ldots,u^T_{ir})&\sim& \mathcal{N}_T(\boldsymbol{0},\Sigma^u_r) \mbox{ with } \Sigma^u_r(t,t') = \tau_r^u f(\kappa^u_{r},t,t'), \nonumber\\
	\epsilon^t_{ij} &\sim ^{\mbox{i.i.d.}}& \mathcal{N}(0, \sigma_e^2). \nonumber
\end{eqnarray}

To model the covariance 
matrices $\Sigma_1^\beta,\ldots,\Sigma_P^\beta$, $\Sigma^a$, and 
$\Sigma_1^u,\ldots,\Sigma_R^u$, we consider two of the commonly used 
covariance functions \citep{rasmussen2004gaussian}, the standard 
exponential, or Ornstein-Uhlenbeck, function and the squared 
exponential function. The standard exponential function has the form
\begin{equation*}
	f(\kappa,t,t') = \mbox{exp}\left(-\frac{|t-t'|}{\kappa}\right),
\end{equation*}
for time points $t$ and $t'$. 
Replacing the distance term by $|t-t'|^2$ gives the 
squared exponential function. When the 
time intervals are moderately spaced, we would prefer the standard 
exponential function. On the other hand, if the time interval is very 
small or smoothness is 
expected, the squared form is preferred.

We assume independent standard normal priors for the diagonal entries of $\mathbf{D}$ for the sake of identifiability. Whereas 
\cite{durante2014} fix the parameter 
$\kappa$ that characterizes the length-scale of the process, in practice we may not have prior 
knowledge about how strongly networks are correlated over time. Thus, we 
choose to estimate $\tau$ and $\kappa$, and the other parameters with the following prior distributions:
\begin{eqnarray*}
\tau^\beta_p &\sim & \mathcal{IG}(k_\beta, \theta_\beta)  \mbox{ and }  
\kappa^\beta_p \sim \mbox{half-Cauchy}(\gamma_\beta) \mbox{ for } 
p=1,\ldots,P,\\
\tau^a &\sim& \mathcal{IG}(k_a, \theta_a) \mbox{ and } \kappa^a \sim 
\mbox{half-Cauchy}(\gamma_a),\\
\tau^u_r &\sim& \mathcal{IG}(k_u, \theta_u) \mbox{ and } \kappa^u_r \sim 
\mbox{half-Cauchy}(\gamma_u) \mbox{ for } r=1,\ldots,R,\\
{d}^t_{r}&\sim^{\mbox{i.i.d.}}&\mathcal{N}(0, 1),\\
\sigma_e^2 &\sim& \mathcal{IG}(k_\sigma, \theta_\sigma).
\end{eqnarray*}

The inverse-Gamma is the most common prior distribution for the variance parameter. The half-Cauchy prior is recommended by \cite{gelman06}.  We experiment different sets of $(k, \theta)$ and $\gamma$, including the g-prior (\cite{zellner86}). The final results are all similar but the mixing performances differ slightly. In the end, we choose $(k, \theta)=(2, 1)$ and $\gamma=5$ so that they are weakly informative with the expected value of the variance parameter being 1. 

All of the model parameters have clear 
interpretation and capture various features of the network. Regression coefficients $\boldsymbol{\beta}$ 
explain the global features as well as observed covariate (nodal and 
dyadic) effects of the network. Additive effects $\boldsymbol{a}$ describe the across-row 
and across-column heterogeneities, which are the same thing in the symmetric case. Its variance $\tau^a$ measures the strength of the dependence induced by sharing the same sender/receiver. Multiplicative factors $\boldsymbol{u}$ capture the non-additive, higher order 
dependence that generates homophily (a social preference to be friends with others similar to you), transitivity (a social preference to be friends with your friends' friends), balance (a social preference to be friends with your enemies' enemies), and stochastic 
equivalence (similar nodes have similar relational patterns, not necessarily connected to each other). 
Non-Gaussian data can be dealt with by 
choosing an appropriate link function in a similar manner as generalized 
linear models.

\subsection{Bayesian Estimation}
\label{sec:PosteriorComputation}
Our posterior computation is performed via a Gibbs sampler to update the 
vector of time-varying regression coefficients and the vector of additive 
and multiplicative latent factors, along with a 
Metropolis-Hastings (MH) algorithm to sample the variance and length GP 
parameters ($\tau, \kappa$).  Full conditional distributions and details about
the MH algorithm are provided in Supplementary Materials Sections A and B (\cite{bomin23}).

Let $\mathbf{E}^t$ denote the $N \times N$ matrix of random 
noise, where the $(i, j)^{th}$ entry is defined as 
\begin{equation*}
			E^t_{ij} = y^t_{ij} - \sum_{p=1}^P 
			X^t_{ijp}\beta^t_{p} - a^t_{i} - a^t_{j} - {\boldsymbol{u}^t_{i}}^\prime 
			\mathbf{D}^t\boldsymbol{u}^t_{j}.
\end{equation*}
Given that the distribution of the observed network $\mathbf{Y}= 
\{\mathbf{Y}^1,\ldots,\mathbf{Y}^T\}$ conditional on all the parameters 
can be written as the product of Normal probability density functions as 
\begin{equation}
			\begin{aligned}
			&\Pr(\mathbf{Y}|\,\mathbf{X}, \boldsymbol{\beta}, \boldsymbol{a}, 
				\boldsymbol{d}, \boldsymbol{u},\sigma_e^2, 
				\boldsymbol\tau^{\beta}, \tau^{a}, \boldsymbol\tau^{d}, 
				\boldsymbol\tau^{u}, \boldsymbol\kappa^\beta, \kappa^a, 
				\boldsymbol\kappa^u)\\&\quad\quad\propto  
				\prod_{t=1}^T\prod_{i>j}(\sigma_e^2)^{-\frac{1}{2}}\mbox{exp}\left\{-\frac{1}{2\sigma_e^2}||E^t_{ij}||^2\right\},
			\end{aligned}
\end{equation}
we sequentially update each parameter from its full conditional 
distribution in the following sampling steps:
		\begin{enumerate}
			\item  Sample $\sigma_e^2 \sim  \mathcal{IG}\big(\frac{T\cdot 
			N(N-1)}{4}+k_\sigma, \frac{1}{2}\sum_{t=1}^T\sum_{i> j}(E^t_{ij})^2 
			+ \theta_\sigma\big)$.
			\item  For each $p = 1,\ldots,P$ in a random order, sample 
			$\boldsymbol{\beta}_{p}$ as follows:
			\begin{enumerate} 
				\item Sample ($\tau^{\beta}_p ,\kappa^\beta_p$) using the 
				MH algorithm of Supplementary Materials Section B (\cite{bomin23}).
			\item Sample $\boldsymbol{\beta}_{p} \sim 
				\mathcal{N}_T\big(\tilde{\mu}_{\beta_p}, 
				\tilde{\Sigma}_{\beta_p} 
				\big)$ with 
				$$\tilde{\Sigma}_{\beta_p} = 
				\Bigg({(\Sigma^\beta_p)}^{-1} + 
				\frac{\mbox{diag}\big(\{\sum_{i>j}{X^{t2}_{ijp}}\}_{t=1}^{T}\big)} 
				{\sigma_e^2}\Bigg)^{-1} 
				\mbox{ and } \tilde{\mu}_{\beta_p} =  
				\Bigg(\frac{\{\sum_{i>j}(E^{t}_{ij[-p]}X^t_{ijp})\}_{t=1}^{T}} 
				{\sigma_e^2}\Bigg)\tilde{\Sigma}_{\beta_p},$$ 
				where $E^{t}_{ij[-p]}=E^t_{ij}+\beta^t_{p}X^{t}_{ijp}$ and 
				$\Sigma^\beta_p=\tau_p^\beta f(\kappa_p^\beta)$.						
			\end{enumerate}
			\item Sample $\boldsymbol{a}_{i}$ as follows:
			\begin{enumerate}
				\item  Sample ($\tau^{a},  \kappa^a$) using the MH 
				algorithm of Supplementary Materials Section B (\cite{bomin23}).
				\item  For each $i= 1,\ldots,N$ in a random order, sample 
				$\boldsymbol{a}_{i} \sim \mathcal{N}_T\big(\tilde{\mu}_{a_i}, 
				\tilde{\Sigma}_{a_i} \big)$ with
				$$\tilde{\Sigma}_{a_i} = 
				\Bigg({(\Sigma_a)}^{-1}+\frac{(N-1)I_T}{\sigma_e^2}\Bigg)^{-1} 
				\mbox{ and }
				\tilde{\mu}_{a_i} = \Bigg(\frac{\{\sum_{ j\neq 
				i}E^{t}_{ij[-i]}\}_{t=1}^{T}}{\sigma_e^2}\Bigg) 
				\tilde{\Sigma}_{a_i},$$ 
				where $E^{t}_{ij[-i]}=E^t_{ij}+a^t_{i}$, $\Sigma^a=\tau^a 
				f(\kappa^a)$.	
			\end{enumerate}
			
			\item For each $r =1,\ldots,R$ in a random order, sample 
			$\boldsymbol{u}_{ir}$ as follows:
			\begin{enumerate}
				\item  Sample  ($\tau_r^{u},  \kappa_r^u$) using the MH 
				algorithm of Supplementary Materials Section B (\cite{bomin23}).
				\item  For $i =1,\ldots,N$ in a random order, sample 
				$\boldsymbol{u}_{ir} \sim 
				\mathcal{N}_T\big(\tilde{\mu}_{u_{ir}}, 
				\tilde{\Sigma}_{u_{ir}} \big)$ with
				$$\tilde{\Sigma}_{u_{ir}} = 
				\Bigg((\Sigma_r^u)^{-1}+\frac{\mbox{diag}\big(\{\sum_{j \neq 
				i}{({d}^t_{r}{{u}^t_{jr}}})^2\}_{t=1}^{T}\big)}{2\sigma_e^2}\Bigg)^{-1}\mbox{ 
				and } \tilde{\mu}_{u_r} =  \Bigg(\frac{\{\sum_{j \neq 
				i}(E^{t}_{ij[-r]}d^t_{r}u^t_{jr})\}_{t=1}^{T}}{2\sigma_e^2}\Bigg) 
				\tilde{\Sigma}_{u_{ir}},$$
				where $E^{t}_{ij[-r]}=E^t_{ij}+{u^t_{ir}}^\prime 
				d^t_{r}u^t_{jr}$ 
				and $\Sigma^u_r=\tau_r^u f(\kappa_r^u)$.
			\end{enumerate}		
		\item For each $r =1,\ldots,R$ in a random order, sample 
		$\boldsymbol{d}_{r}$ as follows:
		\begin{enumerate}
			\item [(a)] Sample $\boldsymbol{d}_{r} \sim 
			\mathcal{N}_T\big(\tilde{\mu}_{d_r}, \tilde{\Sigma}_{d_r} \big)$ 
			with
			$$\tilde{\Sigma}_{d_r} = \Bigg(I_T + \frac{\mbox{diag} 
			\big(\{\sum_{i>j} ({u^t_{ir}u^t_{jr}})^2\}_{t=1}^{T}\big)} 
			{\sigma_e^2}\Bigg)^{-1} \mbox{ and } \tilde{\mu}_{d_r} =  
			\Bigg(\frac{\{\sum_{i>j} 
			(E^{t}_{ij[-r]}u^t_{ir}u^t_{jr})\}_{t=1}^{T}} 
			{\sigma_e^2}\Bigg)\tilde{\Sigma}_{d_r},$$
			where $E^{t}_{ij[-r]}=E^t_{ij}+{u^t_{ir}}^\prime d^t_{r}u^t_{jr}$.
		\end{enumerate}		
		\end{enumerate}
Note that after steps 2 through 5, $\mathbf{E} = 
\{\mathbf{E}^1,\ldots, \mathbf{E}^T\}$ has to be calculated again using the 
previously updated values, so that any update is conditioned on the current 
values of all the other parameters.
		
\subsection{Missing Data}
\label{subsec:VaryingNumberOfNodes}
In dynamic networks,  nodes or edges can be missing at any time point. One 
method to handle such missing data 
is to use the network consisting only of 
the completely observed set of nodes, though we avoid this idea
because it might lead to loss of 
information and potentially biased estimates. 
Another idea \citep[e.g.,][]{sewell2015latent} is to 
treat all missing data as missing at random and impute them. 
%
However, in our application, the specific reason for a missing data point is sometimes
known, as for instance in the case of a nation state that did not exist prior to a certain
year and therefore joined the UN later than other states.  
Moreover, if we have covariates in the data, these covariates might have different
relationships with the missing nodes than with the observed nodes.  
For these reasons, imputation of all missing edge data might generate biases. 
				
To address missing data in our application, therefore, 
we define two types of missing data---``random missing'' and ``structural 
missing''. Let $N$ denote the number of nodes that are part of the network 
at any time from 1 to $T$, then define the $N\times T$ node availability matrix 
$\mathbf{A}$ so that $A_{nt}$ equals 1 or 0, depending on whether or not
node $n$ is available, i.e., non-missing, at time $t$, respectively.
%
Without prior knowledge on the missing mechanism, we 
assume that missing edges corresponding to node $n$ at time $t$ for which 
$A_{nt} = 1$ are random missing, while those for which $A_{nt} = 0$ are 
structural missing. 
Random missing edges are imputed from the current parameter estimates 
at each MCMC iteration, while structural missing edges are excluded from the 
likelihood and the parameters are estimated without them. 
This procedure is beneficial even if the model does not contain covariates, 
since imputing values that are not missing at random could
bias the temporal correlation estimates of additive and/or multiplicative effects.
				
\section{Simulation Study}
\label{sec:simulation study}
Our simulation study has two objectives: First, to show that estimation of 
the correct covariance 
structure plays a key role in the model performance, particularly in the 
case of modeling a network that is highly correlated across time; and second, 
to demonstrate that the eigen-decomposition formulation of the multiplicative 
random effects ($\boldsymbol{u}^\prime \mathbf{D}\boldsymbol{u}$) clarifies 
various transitivity effects.
	
\subsection{Estimating Temporal Correlations} 
\label{subsec:correlation}
We simulate three dynamic network datasets according to 
the generative process in Section~\ref{sec:ExtendingAME}, with $N=20$, $T=10$, $P=1$ (intercept-only), $R=2$, and $(k, \theta) = (2, 1)$ for all $\tau$ parameters as well as $\sigma_e^2$. We use the standard exponential 
covariance function with three different levels of correlation: High ($\kappa=30$), medium 
($\kappa=2$), and low ($\kappa=0.001$).  For each 
generated dataset, we fit the DAME model both while estimating $\kappa$ and  
while holding $\kappa$ fixed at an incorrect value. 

We define the ``lag-1 degree correlation" statistic as the 
Pearson correlation $\rho(\cdot)$ between the vectors 
$v_1 = [\mbox{deg}(\mathbf{Y}^1), \ldots, 
\mbox{deg}(\mathbf{Y}^{T-1})]$ and 
$v_2 = [\mbox{deg}(\mathbf{Y}^2), \ldots, 
\mbox{deg}(\mathbf{Y}^{T})]$, where
%
$\mbox{deg}(\mathbf{Y}^{t})$ is defined as the $N$-dimensional
row vector whose entries are the degree statistics for all nodes.  That is,
$\mbox{deg}(\mathbf{Y}^{t}) = 
(\sum_{j=1}^N y_{1j}^{t}, \sum_{j=1}^N y_{2j}^{t}, \ldots, \sum_{j=1}^N 
y_{Nj}^{t})$.

\begin{figure}[ht]
\centering
\includegraphics[width=0.328\textwidth]{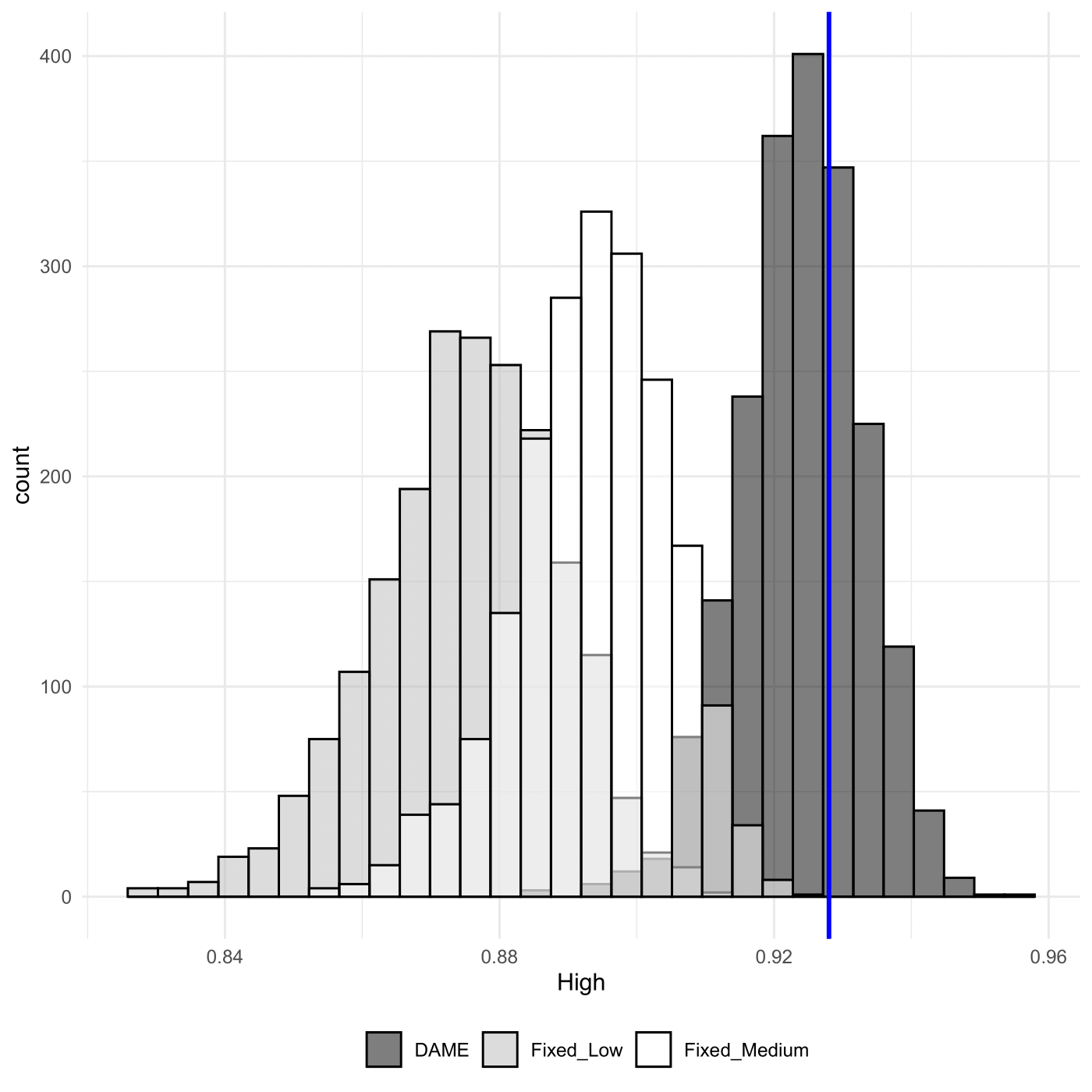}	
\includegraphics[width=0.328\textwidth]{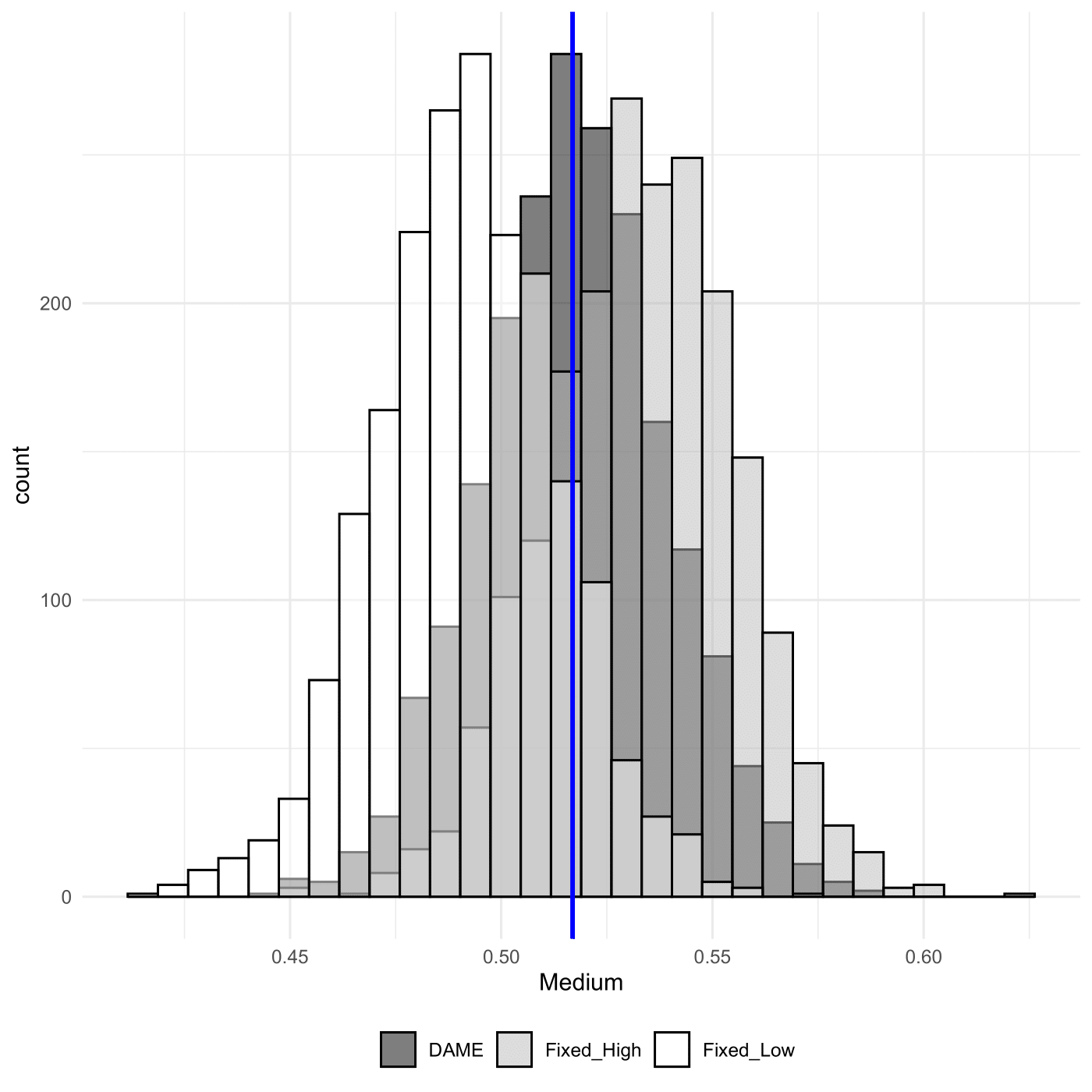}		
\includegraphics[width=0.328\textwidth]{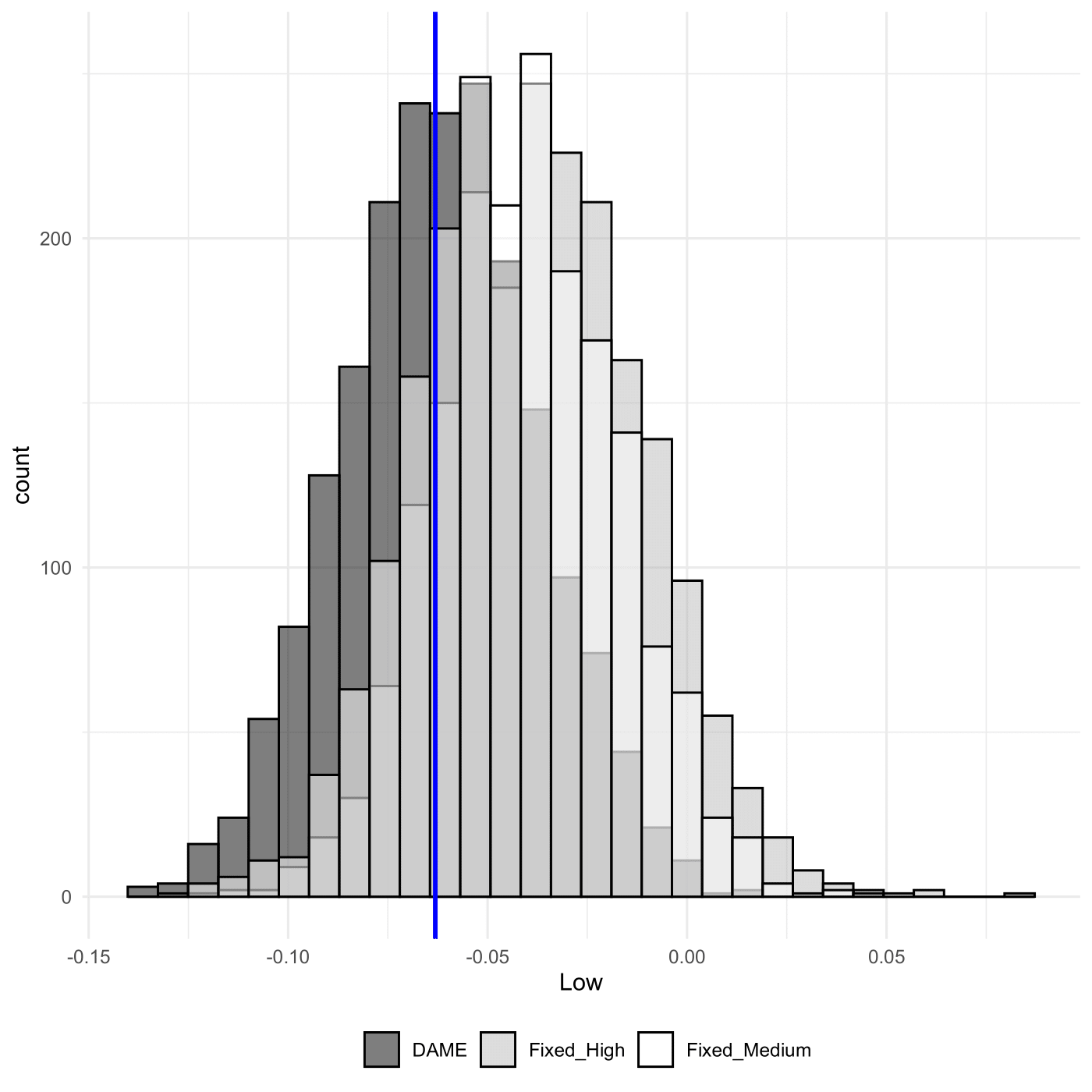}	
\caption {Posterior distributions of lag-1 degree 
correlations. The temporal correlations use the standard exponential
covariance function with $\kappa=30$ (\textit{left}), 
$\kappa=2$ (\textit{middle}), and $\kappa=0.001$ (\textit{right}). The 
vertical line in each plot represents the true value of the lag-1 degree 
correlation.}
\label{figure:correlationstudy}
\end{figure}
		
For all three parameter settings, we run 48,000 MCMC 
iterations, which appears to be enough for convergence, 
calculating the lag-1 degree correlation statistic for a randomly generated 
posterior predicted $\mathbf{Y}$ at each iteration.
In Figure \ref{figure:correlationstudy}, we present the results after discarding the first 8,000
samples and using a thinning interval of 20. For the high correlation case (left), $\kappa$ is 
fixed at 2 and 0.001 to represent misspecifying at medium and low values. 
For the medium correlation case (middle), $\kappa$ is fixed at 30 and 
0.001 to represent misspecifying at high and low values. For the low 
correlation case (right), $\kappa$ is fixed at 30 and 2 to represent 
misspecifying at high and medium values. The true correlation is 
represented by the vertical line in each plot. The plots show
that posterior distributions of the lag-1 degree correlation 
center around the wrong values when $\kappa$ is misspecified, while the full 
DAME model can correctly recover this correlation.  This suggests that in the absence
of prior information about the temporal correlation of the network, it is wise to
estimate $\kappa$ from the data.

\subsection{Estimating Homophily and Transitivity} 
\label{sec:NegativeTransitivity}
As explained in Section~\ref{sec:ExtendingAME},
The $\mathbf{D}^t$ term in Equation~(\ref{eqn:dame}) helps capture both 
homophily and anti-homophily effects.  As explained for example by 
\citet{goodreau2009}, these homophily effects can operate between actors,
in this case nation states, to produce higher order global effects such as transitivity.
We illustrate this phenomenon via a different simulation study.

As in Section~\ref{subsec:correlation}, we take 
$N=20$, $T=10$, $P=1$ (intercept-only), $R=2$, and $(k, \theta) = (2, 1)$ for all $\tau$ parameters as well as $\sigma_e^2$, and simulate model parameters from their respective prior distributions. We use three sets of homophily effects:
First, we take $d_1^t=d_2^t=2$; second, we take $d_1^t=-d_2^t=2$; and third, we take $d_1^t=d_2^t=-2$
for all $t$.
%
%
In each case we fit both the
DAME model as well as the simpler model in which 
${\boldsymbol{u}^t}^\prime \mathbf{D}^t\boldsymbol{u}^t$ is replaced by
${\boldsymbol{u}^t}^\prime \boldsymbol{u}^t$.  Thus, the model is misspecified in the
second and third cases.
%
To isolate the 
high-order effects we wish to illustrate, we assume that the dynamic networks
are independent over time by setting
$(\kappa^\beta, \kappa^a, \kappa^d) = (0, 0, 0)$ and fixing these $\kappa$ parameters
at zero throughout the estimation procedure.

\begin{figure}[!ht]
\centering
\includegraphics[width=0.975\textwidth]{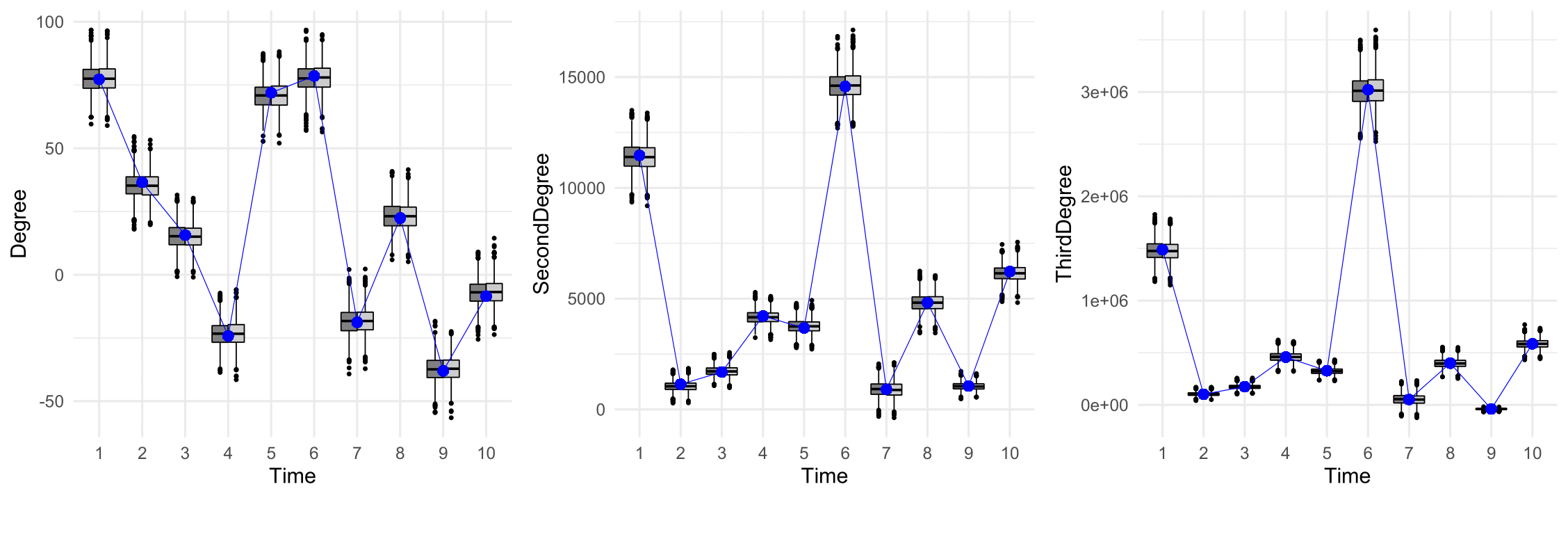}	
\includegraphics[width=0.975\textwidth]{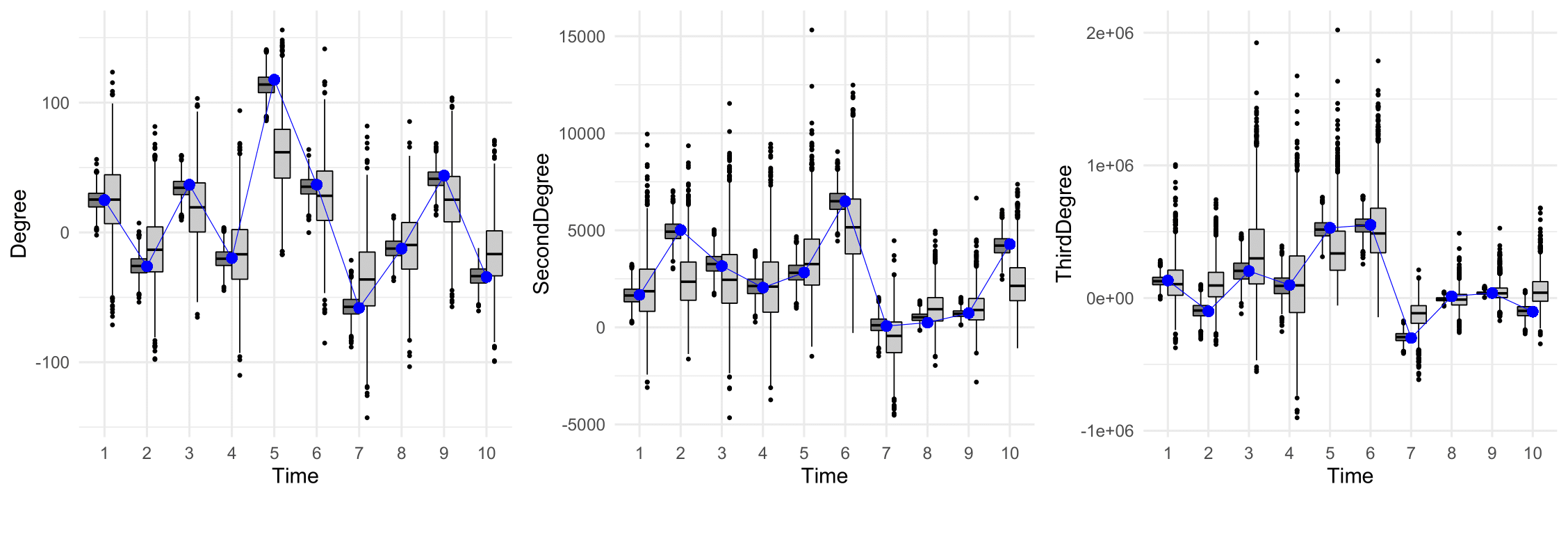}	
\includegraphics[width=0.975\textwidth]{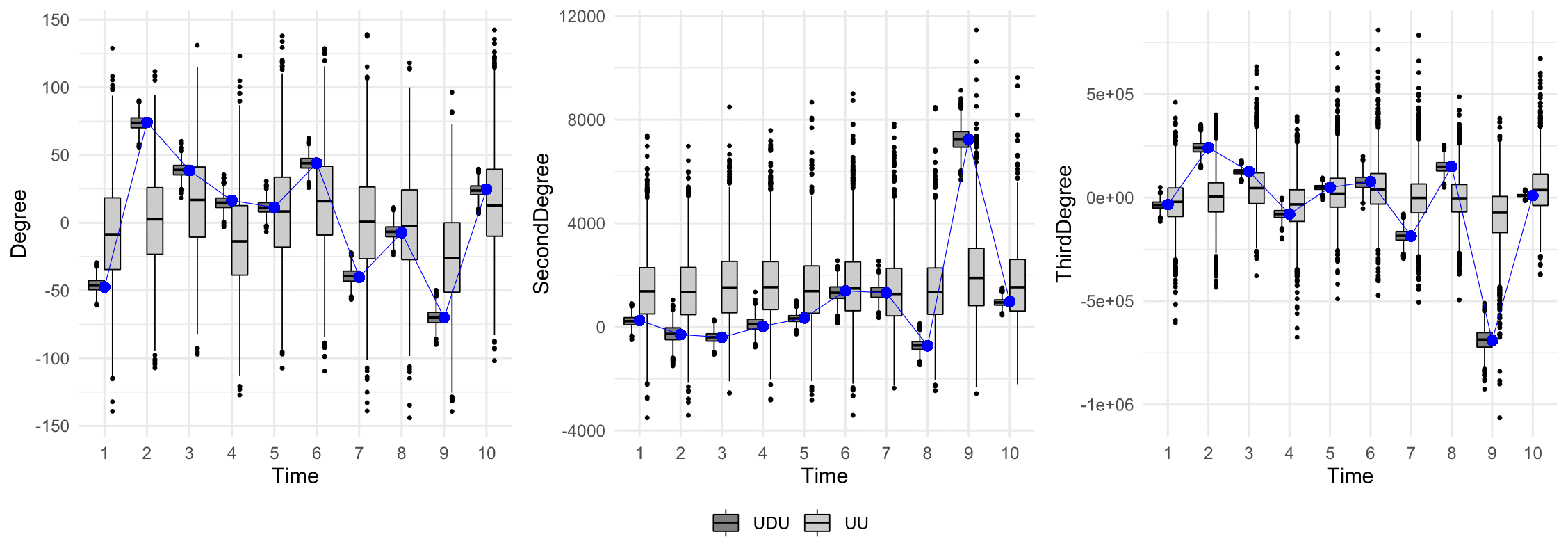}	
\caption {Boxplots of 2,000 posterior predictive degree statistics in the 
first (\textit{left}), second (\textit{center}), and third 
(\textit{right}) moments corresponding to positive (\textit{upper}), mixed 
(\textit{middle}), and negative (\textit{lower}) transitivities: the DAME and the $\boldsymbol{u}^\prime \boldsymbol{u}$ models are shown with the connected dots representing the observed statistics.}
\label{figure:negativitystudy}
\end{figure}

Figure \ref{figure:negativitystudy} gives posterior statistics resulting
from 48,000 MCMC iterations to which we 
apply a thinning interval of 20 after discarding the first 8,000. 
The three columns are the statistics $\mbox{deg$[\mathbf{Y}^t]$}$,
$\mbox{deg$[(\mathbf{Y}^t)^2]$}$, and $\mbox{deg$[(\mathbf{Y}^t)^3]$}$
for a randomly chosen node, where the $N$-dimensional deg operator is 
defined in Section~\ref{subsec:correlation} and 
$(\mathbf{Y}^t)^2$ and $(\mathbf{Y}^t)^3$ are the square and cube
of the matrix $\mathbf{Y}^t$, which represent different orders of moment of the original data and can be viewed as a type of goodness-of-fit statistics.

In the top row, the simpler model using 
${\boldsymbol{u}^t}^\prime \boldsymbol{u}^t$ is correct, and we see no difference 
between the two models' results.  Yet in the case of mixed (second row) or 
negative (third row) transitivity, the two formulations reveal noticeable differences. 
While the DAME model can still recover the true degrees of the first to 
third moments, the simpler model cannot
generate posterior networks that are close to the data in terms 
of the degree statistics. Not only does the simpler 
model introduce bias, but also it 
yields significantly wider interval estimates, implying lower precision 
compared to the DAME model. In addition, the evidence of the 
simpler model's failure to capture the 
transitivity effects becomes larger as the network tends toward stronger 
negative transitivity and also as we move to the degree statistics in 
higher moments. These findings support our choice of the 
DAME framework in order to capture high-order effects such as transitivity.

\section{Analysis of the United Nations Voting Network}
\label{sec:UNvoting}

States' preferences towards each other 
and their stances on different issues help us predict future foreign policies 
and state behaviors, an important topic in the political science literature. 
Votes in the United Nations General Assembly (UNGA) are a standard data source for studying
these preferences, and by considering them as a time-varying network dataset 
using the novel DAME methodology, we hope to address
the following questions:
%
\begin{itemize}
\item What factors are associated with how countries vote? How do those 
associations change over time?
\item Which countries are either more likely  
or less likely to vote with others? How do those behaviors change over time?
\item Beyond observed factors, are there any other unobserved alliances 
that contribute to countries' voting behavior? How do those alliances change 
over time?
\end{itemize}
The dataset and code used in this section can be found at 
\url{https://github.com/bomin8319/DAME}.

\subsection{Data Description}
\label{subsec:DataProcessing}

We consider the time period 1983--2014, which covers important 
international events such as the Cold War and Iraq War. We first determine the 
countries to be included in this analysis by the availability of a set of 
predictors, such as polity score and GDP. We drop countries that have at least 
one predictor with more than 10 years of missing data. Any remaining 
missing values are imputed using the data from previous years. 
The resulting 97 countries are provided along with their abbreviations in Supplementary Materials Section C (\cite{bomin23}).

The voting data are obtained from 
\cite{voeten13}. We use the subset of the votes called ``important votes'', 
identified by the U.S. State Department as ``votes on issues which directly 
affected important United States interests and on which the United States 
lobbied extensively." 
For example, in 2001, important votes include ``Israeli 
Actions in the Occupied Territories,'' ``Peaceful Settlement of the Question of 
Palestine,'' ``U.S. Embargo Against Cuba,'' and ``Nuclear Disarmament.'' Additional 
information on these data is provided by the U.S.~Department of State at
\url{https://www.state.gov/voting-practices-in-the-united-nations/}. The 
number of important votes ranges from 6 to 28 per year, with an average of 12.
Annual voting agreement rates for both important and non-important votes 
are provided in Supplementary Materials Section D (\cite{bomin23}).

The voting data could be viewed as a two-mode, or bipartite, network, where one mode is the set 
of countries and the other is the set of votes.  However, since we are primarily interested
in relationships between countries in this context rather than any features of the votes themselves,
we project the two-mode network to a one-mode network and accept the loss of information that 
this entails.  Each vote has three categories: Y = ``yes'' or approval of an issue, A = 
``abstain,'' and N = ``no'' or disapproval of an issue. As we are interested in the relationship of voting agreement between countries, following \cite{voeten13}, we define an indicator variable $y_{ijt}^k$ as agreement
between each pair of countries on each vote in which they both participate, and define abstain with others as half agreement, i.e.
\[
y_{ijt}^k = 
\begin{cases}
1 & \mbox{if $i$ and $j$ vote Y/Y, N/N, or A/A} \\ 
0.5 & \mbox{if $i$ and $j$ vote Y/A or N/A} \\
0 & \mbox{otherwise}.
\end{cases}
\]
Each vote is on a specific matter, and the relational network constructed from the vote-by-vote data would give us estimates of countries' voting patterns that jump around reflecting the UN agenda more than countries' stand. Therefore, instead of the vote-specific agreement, we take the annual average agreement to be our response. Specifically, let
$n_{ij}^t$ denote the number of votes in which both $i$ and $j$ participated
in year $t$ and define $Y_{ij}^t=(\sum_k y_{ijt}^k / n_{ij}^t )\times 100$. The combined response $\mathbf{Y} = \{\mathbf{Y} ^1,\ldots,\mathbf{Y} 
^{32}\}$ is a $97 \times 97 \times 32$-dimensional array representing the percentage of annual agreement and taking values between 
0 and 100. 

Next, we construct $P=6$ covariates 
from the Correlates of War data \citep{gibler2008international}, the Polity IV data 
\citep{marshall2014polity}, the International Monetary Fund's Direction of 
Trade Statistics, and the International Financial Statistics data. For 
$p=1,\ldots,P$, the explanatory variable $X^t_{ijp}$ is the following:
\begin{enumerate}
			\item  $X^t_{ij1}$: The constant 1, an intercept to account for the 
			baseline degree of agreement.
			\item  $X^t_{ij2}$: distance from gravity model (\cite{ward2007persistent}): log(geographic distance between the 
			capital cities of country $i$ and country $j$) .
			\item  $X^t_{ij3}$: Indicator of whether country $i$ and country $j$ have a 
			formal alliance including mutual defense pacts, non-aggression 
			treaties, and ententes at time $t$.
			\item  $X^t_{ij4}$: absolute difference in Polity IV score between 
			country $i$ and country $j$ at time $t$. These scores range from 
			$-10$ to $+10$ and 
			provide annual information on the level of democracy, with 
			$-10$ to $-6$ corresponding to autocracy, $-5$ to 5  
			to anocracy, and 6 to 10 to democracy.
			\item  $X^t_{ij5}$: index of economic dependence using bilateral 
			trade weighted by each country's gross domestic product (GDP), as 
			defined in \cite{gartzke2000preferences}. That is, $$X^t_{ij5} = 
			\min\Big(\frac{\mbox{Trade}_{ijt}}{\mbox{GDP}_{it}}, 
			\frac{\mbox{Trade}_{ijt}}{\mbox{GDP}_{jt}}\Big).$$
			\item  $X^t_{ij6}$: indicator of whether country $i$ and country 
			$j$ share the official language.
\end{enumerate}
By definition, all covariates are symmetric, i.e., $X^t_{ijp}=X^t_{jip}$, and all but two
of them, $\log(\mbox{distance})$ and common language, may vary over time.
Correlations between the covariates are summarized in 
Supplementary Materials Section D (\cite{bomin23}).

The matrix of availability $\mathbf{A}$, introduced in 
Section~\ref{subsec:VaryingNumberOfNodes}, reflects some countries'
non-participation in the United Nations General Assembly (UNGA), as follows:
\begin{itemize}
	\item  North Korea (PRK) and South Korea (ROK) have 
	structural missing values from $t = 1$ to 
	$t = 8$ because neither voted until North Korea and South Korea 
	were simultaneously admitted to the United Nations in 1991.
	\item Russia (RUS) has structural missing values from $t=1$ to $t=9$ 
	because Russia assumed the Soviet Union's seat, including its permanent 
	membership on the Security Council in the United Nations, after the 
	dissolution of the Soviet Union in 1991. 
	\item Iraq (IRQ) has structural missing values from $t =13$ to $t = 21$ 
	because it did not participate in UNGA roll call votes from 1995 to 
	2003. Under the rule of Saddam Hussein, Iraq had been under sanctions from 
	the international community, including the UN, since 1990.
\end{itemize}
As explained in
Section~\ref{subsec:VaryingNumberOfNodes}, other missing values are treated
as missing at random and thus imputed.
\subsection{Model Validation}
\label{subsec:Model Validation}

\begin{figure}[!ht]
 	\begin{center}
 		\includegraphics[width=0.775\textwidth]{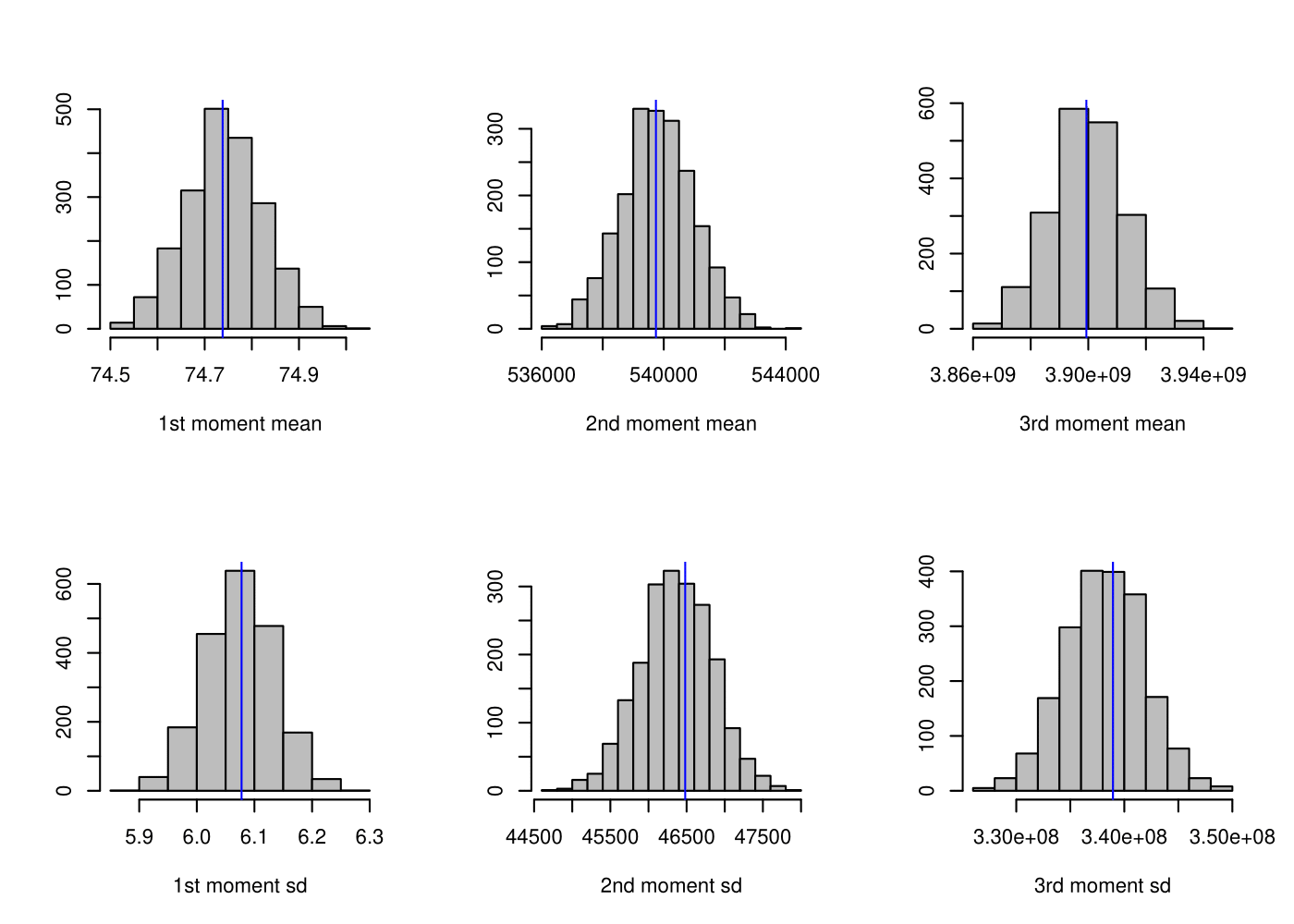}
 	\end{center}
 	\caption {Posterior predictive distribution of the mean and standard 
 	deviations of first, second, and third order moment statistics. The observed 
 	statistics are represented by the vertical lines.}
 	\label{figure:modelvalidation1}
 \end{figure}

Before interpreting the results of fitting the DAME model, we generate
multiple posterior predictive datasets from the fitted model for the purpose of validation. 
Year by year, we compare the mean and standard deviation of the first-, 
second-, and third-order moment statistics, as defined in Section~\ref{sec:NegativeTransitivity}, 
of the generated datasets to the observed statistics. 
Figure~\ref{figure:modelvalidation1} shows that the 
model-generated statistics resemble those features of the observed data well. 
We use $R=2$ for the dimension of the latent space of the multiplicative effects in 
model~(\ref{eqn:dame}).  This choice is based on informally experimenting with
several possible choices for $R$ rather than a formal model selection procedure;
we say a bit more about this choice in Section~\ref{sec:Discussion}.

\begin{figure}[!ht]
\begin{center}
\includegraphics[width=1.0\textwidth]{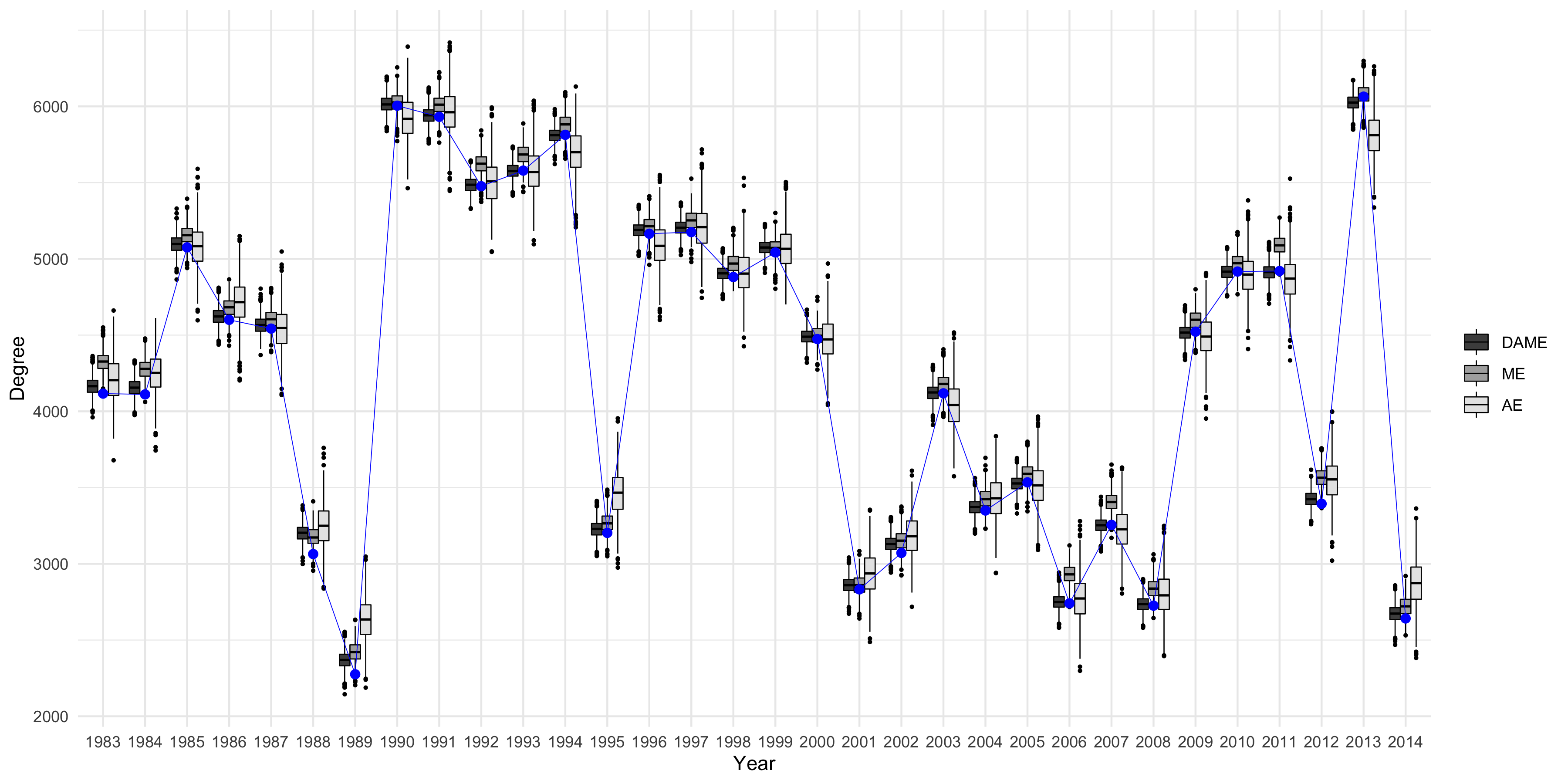}	
\end{center}
\caption {Boxplots of 1,000 posterior predictive degree statistics for the 
United States (USA) for three models: The full DAME with both additive and multiplicative effects, the multiplicative effects only model ME, and the additive effects
only model AE. The connected dots show
the observed statistics.}
\label{figure:modelvalidation}
\end{figure}

Finally, we investigate the necessity of including each component in the DAME
model.  Specifically, we fit four different special cases of the model~(\ref{eqn:dame}), 
according to whether the additive effects $a_i^t + a_j^t$ and/or the multiplicative 
effects $\boldsymbol{u}^t_{i}\mathbf{D}^{t} \boldsymbol{u}^t_{j}$ are included.  
Each of the four cases includes the covariate term $X^t_{ijp}\beta^t_{p}$.
Figure~\ref{figure:modelvalidation} depicts the degree statistics constructed from 
1,000 different posterior predictive samples, where the degrees are normalized 
for better visibility. Here we take the United States 
(USA) as an example to discuss the general findings. An alternative figure with the model NO with neither effects are provided in Supplementary Materials Section E (\cite{bomin23}).

Figure~\ref{figure:modelvalidation} shows that inclusion of either the additive effects 
or the multiplicative effects has the effect of reducing bias, and the AE model has slightly smaller
bias than the ME model.  Furthermore, the multiplicative effects evidently increase the precision of 
the estimates. In other words, it appears that the use of both additive and multiplicative effects
in the full DAME model achieves both low bias and high precision.

Calculating the sum of squared errors (SSE) of the degree statistics over all 1,000 posterior samples, 
97 countries, and 32 time points for each of the four models of Figure~\ref{figure:modelvalidation}, where
\[
\mbox{SSE} = \sum_{t=1}^{32} \sum_{k=1}^{1000} 
\left\| \text{deg}(\mathbf{Y}^t) - \text{deg}(\mathbf{\hat{Y}_k}^t) \right\|^2,
\] 
we obtain 5.539 for the NO model, 0.148
for AE, 0.035 for ME , and 0.018 for DAME.

In addition, we perform simulation studies to demonstrate the robustness of our model under violations of the normality assumption (Supplementary Materials Section F (\cite{bomin23})). Goodness of fit plots which evaluate how well the models fit the observed networks based on degree statistics and transitivity property \citep{hunter2008goodness} are provided in Supplementary Materials Section G (\cite{bomin23}).

\subsection{Results and Interpretation}
\label{subsec:UNresult}

The results here are based on $150{,}000$ Gibbs 
iterations. Due to strong autocorrelation among the parameters, we tuned the 
chain with a burn-in of $50{,}000$ and thinned the iterations by keeping every 
$50^{\rm {th}}$ sample. 

\begin{figure}[ht]
\begin{center}
\includegraphics[width=1.0\textwidth]{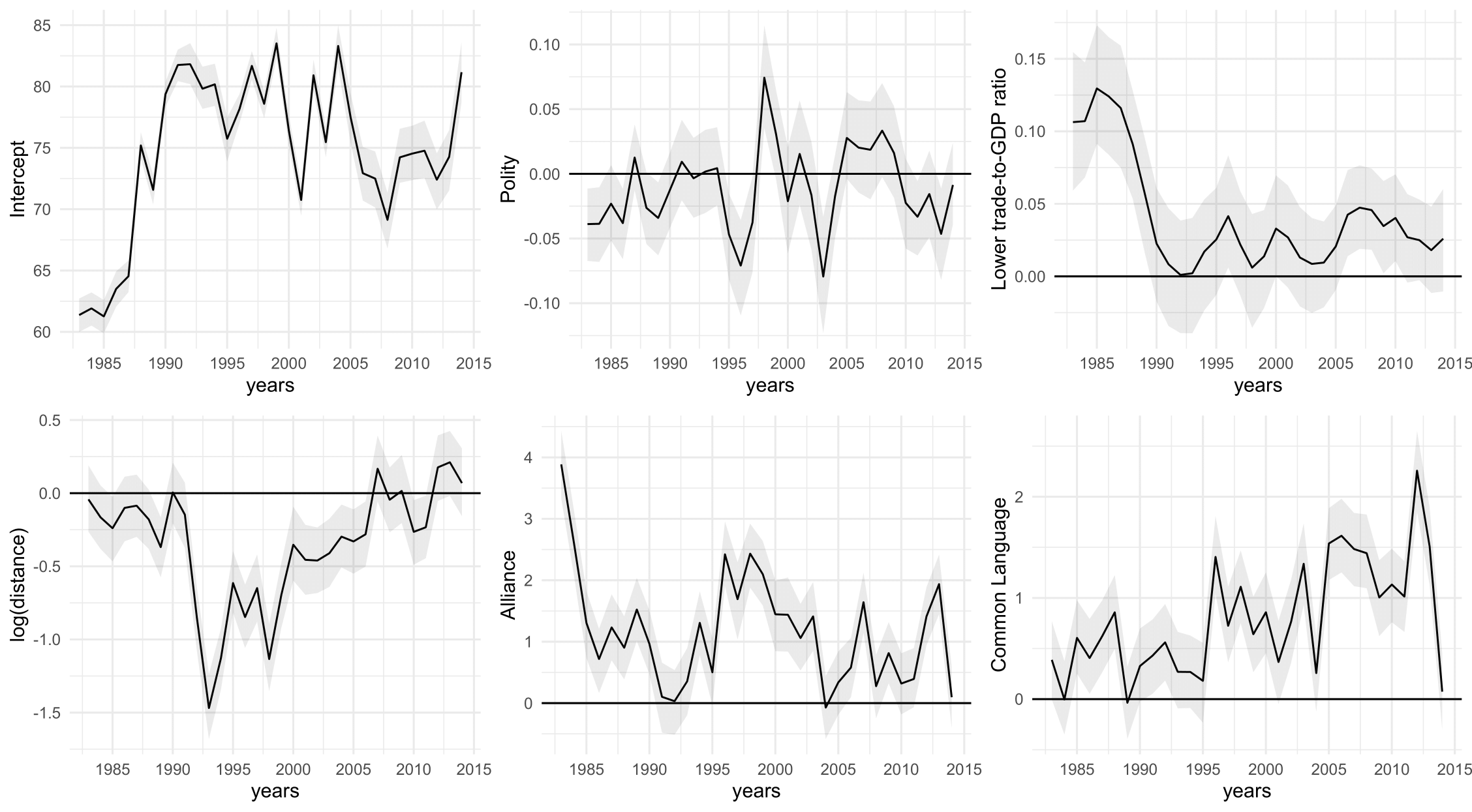}
\end{center}
\caption {Posterior mean estimates (colored lines) and
95\% credible intervals (gray areas) for the fixed effect coefficients 
$\boldsymbol{\beta}$ corresponding to intercept, log(distance), 
alliance, polity difference, lower trade-to-GDP ratio, and common language.}
\label{figure:interceptplot}
\end{figure}

Figure \ref{figure:interceptplot} shows the posterior mean estimates of the 
fixed effects coefficients $\{\boldsymbol{\beta}_p\}_{p=1}^P$ with their 
corresponding 95\% credible intervals. Overall, the effects of the covariates on 
the United Nations voting behavior change substantially over time, especially in 
the cases of geographic distance and trade-to-GDP ratio. Most importantly, the 
critical juncture for these temporal changes seems to be around the end of 
Cold War, that is, the late 1980s and early 1990s. For instance, the middle 
panel of the top row reveals the pattern of influence of geographic distance on 
voting behavior. The gravity model \citep{leibenstein1966shaping, 
rodrigue2009geography} suggests that bilateral flows, e.g., trade, migration, 
or portfolio investment, between two countries are a negative 
function of the distance between them. We see an overall negative coefficient 
for geographic distance, which is consistent with the gravity model. However, the 
negative effect of geographic distance is less significant after the early 
1990s. It is likely that the votes in the United Nations were much more 
influenced by the overall ideological conflicts between the Soviet Union plus 
its satellites on one side and the Western camp on the other, 
so the effect of geographical distance was 
weakened during the Cold War.

Regarding the effect of polity---or 
distance between polity scores as we operationalize this variable---the result suggests that 
in general, political regime similarity is not often associated with high 
agreement in the United Nations General Assembly, at least for a few time 
periods included in the study, e.g., 1990--1995, 1997--2002, and 2005--2010. 
Scholars in the liberal tradition of international relations have long argued 
for shared norms, values, and preferences between democracies; what the 
result here suggests is that such similarity in preferences is not 
sufficiently strong to sway countries' votes in the United Nations 
General Assembly, at least in the case of important votes and when we account 
for the other factors in the model. On the other hand, as we expect, having alliances, 
active trade, and common languages all have positive effects on voting
similarity. The posterior mean estimates of $\kappa$'s, the range parameters of Gaussian process, corresponding to the fixed effects are $\kappa_{intercept}^\beta = 68.514, \kappa_{log(distance)}^\beta = 4.817, \kappa_{polity}^\beta = 194.444, \kappa_{alliance}^\beta = 3.587, \kappa_{trade/GDP}^\beta = 650.589,$ and $\kappa_{language}^\beta = 3.085$, which show that these time-varying parameters are correlated over time with long memories.

\begin{figure}[!h]
\begin{center}
\includegraphics[width=1\textwidth]{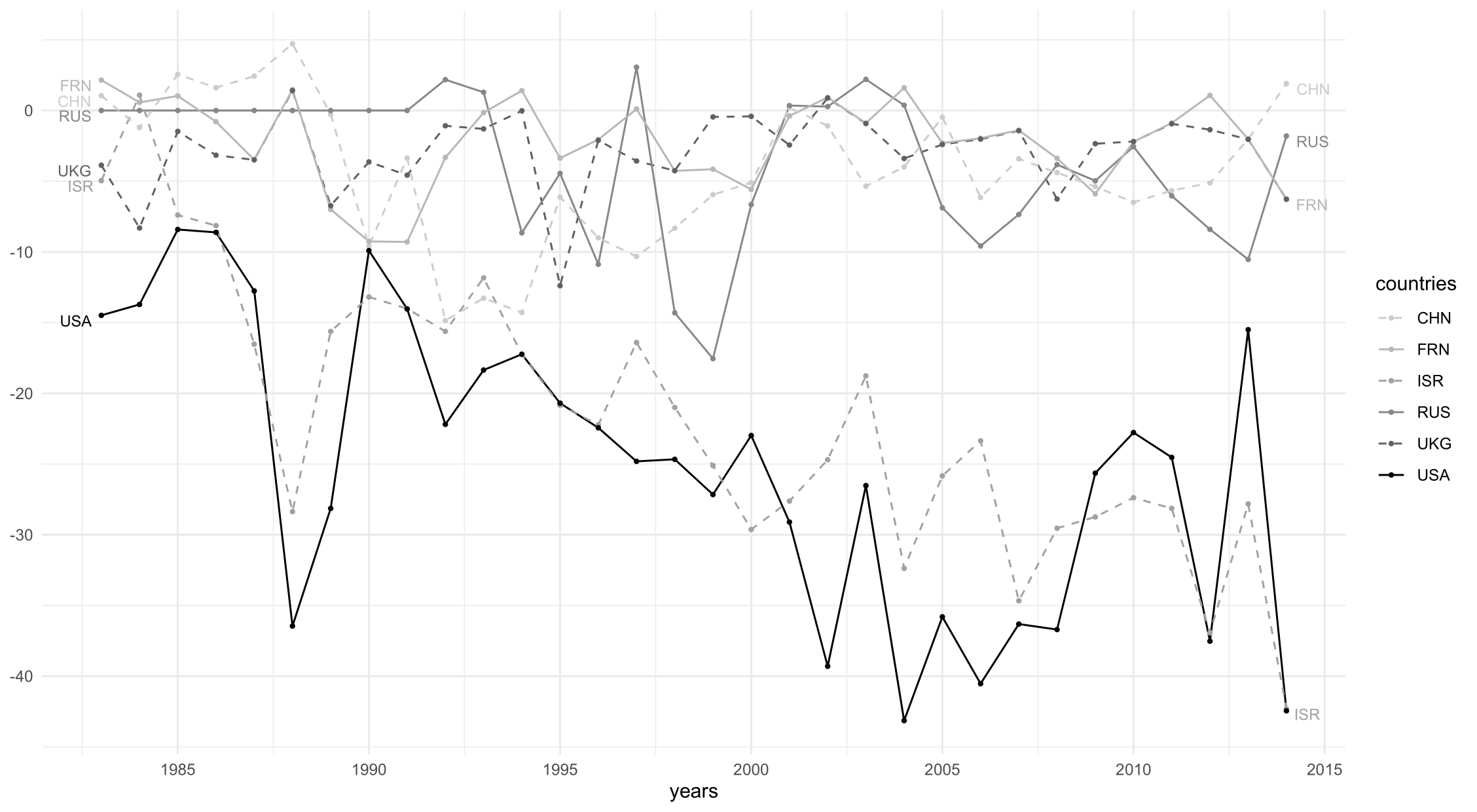}
\end{center}
\caption {Posterior mean estimates for the additive random effect estimates 
$\boldsymbol{a}^t$.}
\label{figure:thetaplot}
\end{figure}

As for the random effects, for clear visualization we only present the results from 6 countries where 5 are permanent members (China, France, Russia, the United Kingdom, and the United States) and Israel being added based on its large additive random effects. These 6 countries are marked with asterisks in Supplementary Materials Section C (\cite{bomin23}).
Figure~\ref{figure:thetaplot} shows the posterior mean estimates of each country's 
additive random effects, that is, its node-specific time-varying intercepts 
$a_i^t$ from Model~(\ref{eqn:dame}). 
Here, the United States (USA) and Israel (ISR) stand out with 
large negative additive random effects, suggesting that these two countries are 
less likely to vote in agreement with other countries.  Considering 
that the majority of votes are ``yes", this implies among other things
that USA and ISR are more likely to vote ``no" than other countries. The posterior mean of the range parameter $\kappa$ for the additive effects is a 4.001, indicating a relatively strong memory up to lag 3 or lag 4.  

\begin{figure}[!h]
\begin{center}
\includegraphics[width=1\textwidth]{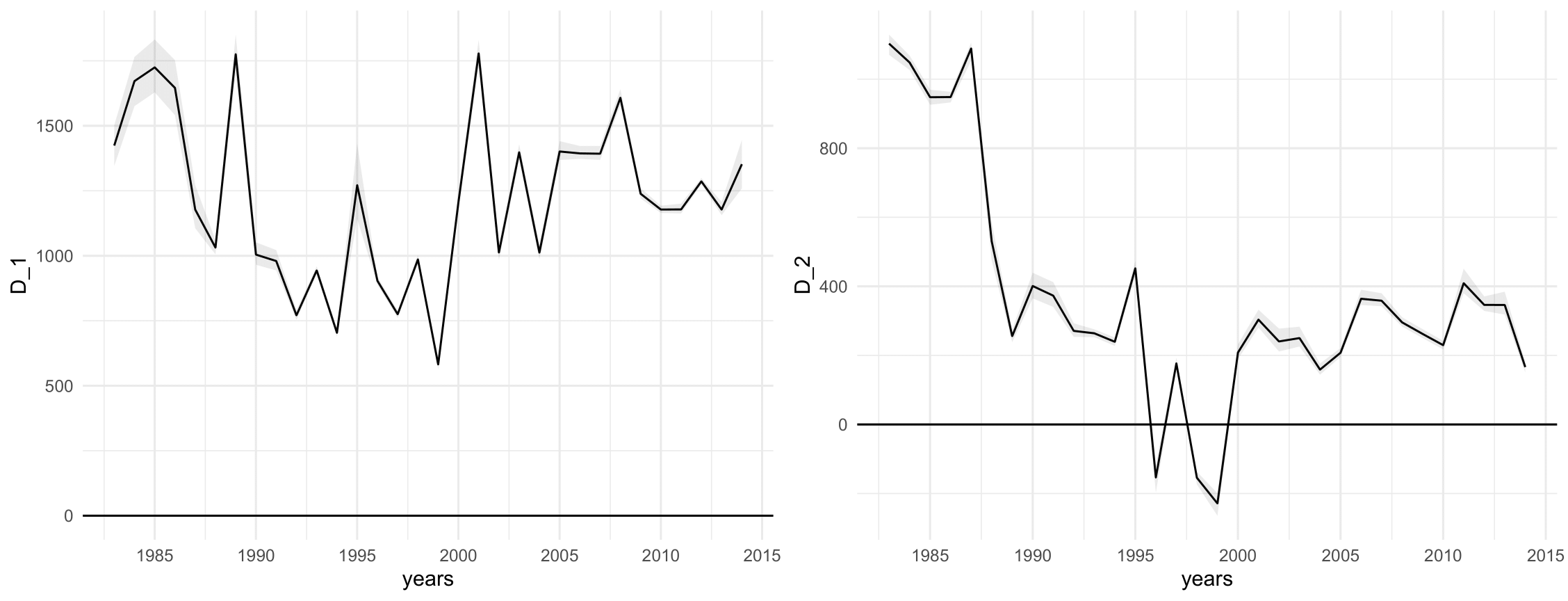}	
\end{center}
\caption {Posterior mean estimates and 95\% credible intervals for the eigenvalues 
$d_1^t$ and $d_2^t$ from the eigen-decompositions of 
${\boldsymbol{u}^t}^\prime 
\mathbf{D}^t\boldsymbol{u}^t$ in the 32 years from $t=1983$
to $t=2014$.}
\label{figure:Dplot}
\end{figure}

Since the DAME model~(\ref{eqn:dame}) depends on $\boldsymbol{u}$ and
$\mathbf{D}$ only through $\boldsymbol{u}^\prime 
\mathbf{D}\boldsymbol{u}$, to preserve identifiability we calculate an
eigen-decomposition on every posterior 
matrix $\boldsymbol{u}^\prime 
\mathbf{D}\boldsymbol{u}$, then take $\mathbf{D}$ to be the diagonal matrix of 
eigenvalues and $\boldsymbol{u}$ the corresponding matrix of eigenvectors. 
Figure~\ref{figure:Dplot} shows that most eigenvalues are positive, implying general tendency of positive homophily in the countries' latent positions.  

In year 1996, 1998, and 1999, the eigenvalues in the second dimension (r=2) are negative, implying the possible existence of anti-homophily. The posterior mean of the range parameter $\kappa$ for the first and second dimension of $\boldsymbol{u}$ is  10.874 and 0.867, indicating the first dimension of the latent position has a strong and long memory while the second dimension has a short memory similar to Markovian one.

\begin{figure}[!ht]
\begin{center}  
\includegraphics[width=1\textwidth]{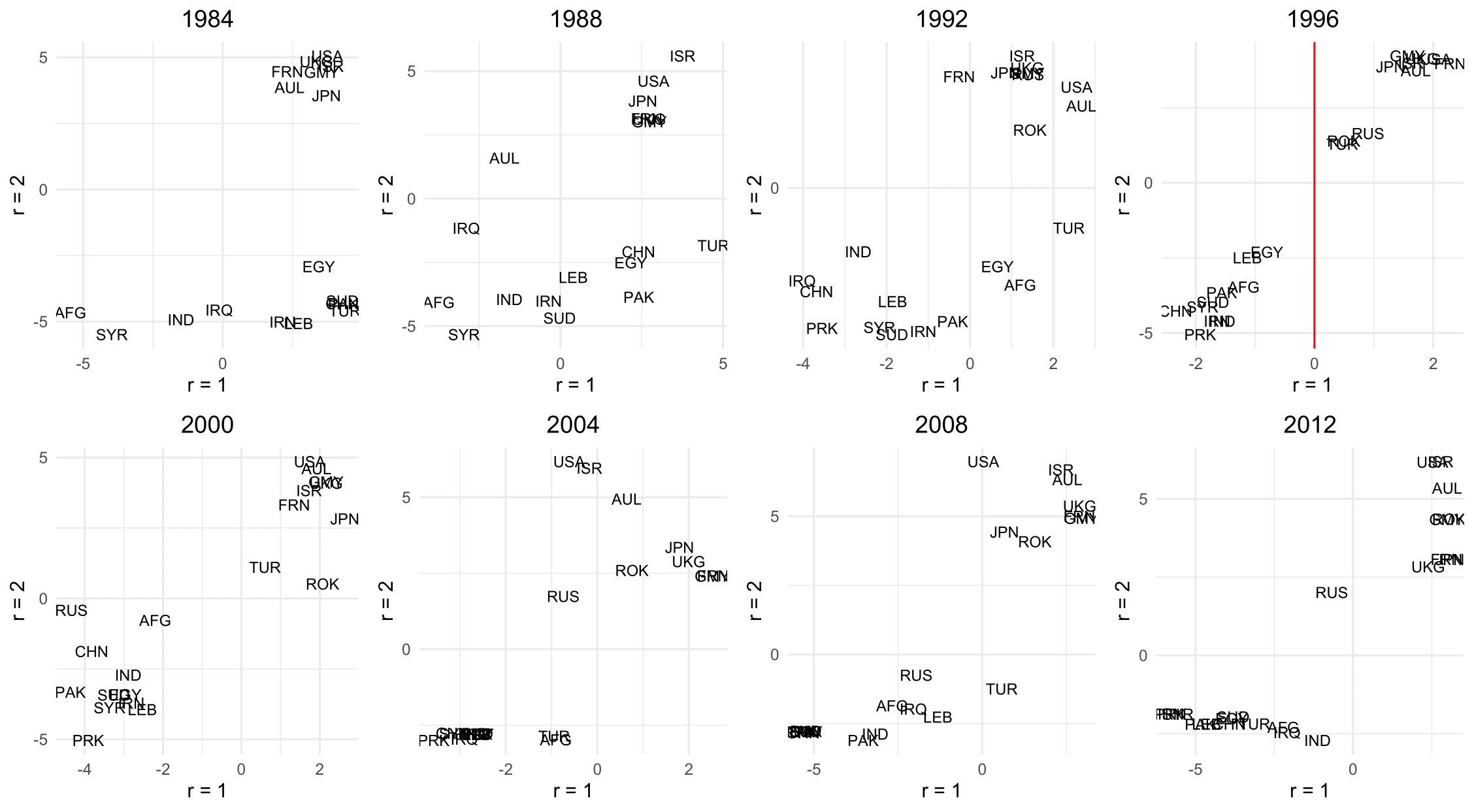}
\end{center}
\caption {Posterior mean estimates of the multiplicative random effects 
$\boldsymbol{u}_i^t{\sqrt{\mathbf{D}^t}}$ for 21 selected countries for 
every 4 years between $t=1984$ and $t=2012$. For 1996, the solid line of the y-axis indicates a negative $d_2$ value. }
\label{figure:UDplot}
\end{figure}

To visualize the latent multiplicative-effect positions,
we apply a Procrustes transformation of each $\boldsymbol{u}^t\sqrt{\mathbf{D}^t}$ with $\boldsymbol{u}^{t-1}\sqrt{\mathbf{D}^{t-1}}$ as the target matrix. In Figure \ref{figure:UDplot}, we plotted the scaled latent positions $u_1^t \sqrt{d_1^t}$ vs $u_2^t \sqrt{|d_2^t|}$. When $d_1^t > 0$ and $d_2^t > 0$, countries close together and far from the origin have higher propensity to agree in their UN voting patterns. When $d_2^t<0$, such as year 1996 (as shown in Figure \ref{figure:UDplot} with y-axis highlighted in red), countries close in their x-axis and far away in their y-axis have higher propensity to agree in their UN voting patterns.

We see interesting 
patterns of clustering of countries over time. Over the years, the most consistent pattern we observe from the latent space is a cluster of the USA and its close allies. For example, in 1984 – this is the later stage of the Cold War, US (USA, higher right hand side of the plot) is clustered with a number of countries, all of them are close US allies: United Kingdom (UKG), Israel (ISR), West Germany (GMY), Japan (JPN), France (FRN), and Australia (AUL). On the other side of the latent space locate mostly the non-US-ally countries such as Egypt (EGY), Iraq (IRQ), and Syria (SYR). Four years later, in 1988, this is a few years before the end of the Cold War, the 1984 pattern persisted; the US cluster we observed in 1988 almost stayed the same with the only exception being Australia (AUL) that moved to mid-way between the US cluster the rest of the countries in the latent space. However, in 1992, Australia (AUL) moved back to the US cluster which now also included Korea Republic (ROK) and Russia (RUS). Russia’s location in this year reflects the Honeymoon period between the newly independent Russia and the United States right after the collapse of the Soviet Union. Indeed, after the end of the Cold War, diplomatic relations between the US and Russia improved until the end of the late 1990s. For example, both nations signed an arms control treaty in early 1993; they bilateral trade also increased. However, the US-Russia relations began to deteriorate in the late 1990s, starting in 1997 when NATO offered membership to Hungary, Poland, and the Czech Republic. Russia’s position in the latent space reflects such a change in relationship. For example, in 2000. Russia (RUS) has moved to the opposite side from the US cluster. In all later years shown in the plots, 2004, 2008, and 2012, Russia never rejoined the US cluster, which reflects the tensions between the two countries ever since the NATO expansion in the late 1990s. We see that in 1992 and 2000, countries lie in two clusters and they vote in great agreement within their cluster. This implies high transitivity. On the other hand, in 1996, countries' voting behaviors are similar within a cluster but their votes do not necessarily agree with those within their cluster. This demonstrates stochastic equivalence. Our model captures both of these network features through the multiplicative effects $u$ and the sign of $D$.

\section{Discussion}
\label{sec:Discussion}

As an extension of the additive and multiplicative effects (AME) model, the 
dynamic additive and multiplicative effects (DAME) network model can flexibly 
learn the underlying time-varying structure in dynamic networks, while inferring 
the effects of node-specific and dyad-specific latent variables. Accounting for 
the correlation structure of the networks makes better use of dynamic networks 
than modeling them as separate network snapshots. The visualization of the 
model-estimated time-varying parameters provides an effective temporal trend 
analysis of dynamic networks, as well as the descriptive visualization of 
higher-order dependence over time. In modeling UN voting 
behavior, the example that has motivated our model, we find that having alliances, active trade, and common languages are positively associated with countries' voting similarities. The nodal effects indicate that USA and Israel are generally less likely to vote in agreement with other countries. After controlling for the known factors in the studies of international relations and nodal heterogeneity, the multiplicative effects reveal interesting clustering patterns in the latent space, such as the allies of USA, and their position changes relative to the Soviet Union and Russia, before and after the Cold War.  
These particular data display both positive and negative homophily over time which showcase the advantage of the model, however, the methodology proposed here goes beyond this 
particular example and can be applied to a wide range of applications. 

Several extensions and modifications of this work are possible.
If researchers are 
interested in the determining factors in changes of voting behaviors, we could 
include $Y_{t-1}$ in the model, as in \cite{friel2016interlocking}. Given the 
evidence found in the data that the current year's outcome highly depends on the 
previous year, this additional term could also improve the goodness of fit of 
the model. In addition, our model assumes a common variance for all additive effects. With the added time dimension, more complicated models with heterogeneous additive effects are possible. 

Although we illustrate the entire framework in the context of symmetric or 
undirected networks, it is straightforward to apply the same dynamic extension 
to the directed version of the additive and multiplicative effects model of
\citet{hoff2021additive}.  Furthermore, our framework can be applied to 
binary and ordinal network data with appropriate link functions. 

We have avoided any sort of formal model-selection criterion in choosing the 
value of $R$, the dimension of the latent multiplicative effect space.  For the
UNGA voting data, we settled on $R=2$ based on informal exploration
suggesting that increasing the dimension does not markedly improve the model fit. 
For instance, the estimated posterior eigenvalue $d^t_{3}$ was roughly zero for all 
$t=1,\ldots,T$.  There exists the possibility of developing a more rigorous method
of selecting this dimension.

Finally, considering the increasing ubiquity of large network datasets with 
time granularity, future work might focus on 
improving the computational efficiency of the fitting algorithm.
Our current MCMC algorithm mixes slowly, taking 
over 72 hours to achieve full convergence of all fixed and random effects for the UNGA dataset. Surely this performance 
could be improved.

\begin{supplement}
\stitle{A. Derivations of Full Conditional Distributions}
\sdescription{This section provides the derivations of the full conditional distributions for estimating the model parameters.}
\end{supplement}
\begin{supplement}
\stitle{B. Metropolis-Hastings Algorithm for GP Parameters}
\sdescription{This section describes the Metropolis-Hastings Algorithms for estimating the Gaussian Process parameters in the model.}
\end{supplement}
\begin{supplement}
\stitle{C. List of Countries in Voting Network}
\sdescription{This section provides the full list of countries in the UN voting network.}
\end{supplement}
\begin{supplement}
\stitle{D. Summary Statistics }
\sdescription{This section presents exploratory data analysis statistics of the dynamic voting data, including summary statistics of the total votes and important votes, and associations among the predictors used in the model.}
\end{supplement}
\begin{supplement}
\stitle{E. Posterior Predictive Statistics for the United States (USA) }
\sdescription{This section presents boxplots of posterior predictive degree statistics of four models U.S.A.}
\end{supplement}
\begin{supplement}
\stitle{F. Residual Diagnostics}
\sdescription{This section presents the residual plots and additional simulations to demonstrate the robustness of the model to mis-specification of the error distribution. }
\end{supplement}
\begin{supplement}
\stitle{G. Posterior Summaries on GP Parameters }
\sdescription{This section presents tables that describe the posterior distributions of the Gaussian Process parameters.}
\end{supplement}
\begin{supplement}
\stitle{H. Posterior Predictive Checks}
\sdescription{This section presents posterior predictive plots of the overall degree distributions aggregating all nodes and
time points.}
\end{supplement}

\bibliographystyle{chicago}
\bibliography{BominBib2}

\begin{thebibliography}{}

\bibitem[\protect\citeauthoryear{Bailey, Strezhnev, and Voeten}{Bailey
  et~al.}{2017}]{bailey2017estimating}
Bailey, M.~A., A.~Strezhnev, and E.~Voeten (2017).
\newblock Estimating dynamic state preferences from {U}nited {N}ations voting
  data.
\newblock {\em Journal of Conflict Resolution\/}~{\em 61\/}(2), 430--456.

\bibitem[\protect\citeauthoryear{Durante and Dunson}{Durante and
  Dunson}{2014}]{durante2014}
Durante, D. and D.~B. Dunson (2014).
\newblock Nonparametric {B}ayes dynamic modelling of relational data.
\newblock {\em Biometrika\/}~{\em 101\/}(4), 883--898.

\bibitem[\protect\citeauthoryear{Friel, Rastelli, Wyse, and Raftery}{Friel
  et~al.}{2016}]{friel2016interlocking}
Friel, N., R.~Rastelli, J.~Wyse, and A.~E. Raftery (2016).
\newblock Interlocking directorates in {I}rish companies using a latent space
  model for bipartite networks.
\newblock {\em Proceedings of the National Academy of Sciences\/}~{\em
  113\/}(24), 6629--6634.

\bibitem[\protect\citeauthoryear{Gartzke}{Gartzke}{1998}]{gartzke1998kant}
Gartzke, E. (1998).
\newblock {K}ant we all just get along? {O}pportunity, willingness, and the
  origins of the democratic peace.
\newblock {\em American Journal of Political Science\/}~{\em 42}, 1--27.

\bibitem[\protect\citeauthoryear{Gartzke}{Gartzke}{2000}]{gartzke2000preferences}
Gartzke, E. (2000).
\newblock Preferences and the democratic peace.
\newblock {\em International Studies Quarterly\/}~{\em 44\/}(2), 191--212.

\bibitem[\protect\citeauthoryear{Gelman}{Gelman}{2006}]{gelman06}
Gelman, A. (2006).
\newblock Prior distributions for variance parameters in hierarchical models
  (comment on article by browne and draper).
\newblock {\em Bayesian Analysis\/}~{\em 1\/}(3), 515--534.

\bibitem[\protect\citeauthoryear{Gibler}{Gibler}{2008}]{gibler2008international}
Gibler, D.~M. (2008).
\newblock {\em International military alliances, 1648-2008}.
\newblock CQ Press.

\bibitem[\protect\citeauthoryear{Goodreau, Kitts, and Morris}{Goodreau
  et~al.}{2009}]{goodreau2009}
Goodreau, S.~M., J.~A. Kitts, and M.~Morris (2009).
\newblock Birds of a feather, or friend of a friend? using exponential random
  graph models to investigate adolescent social networks.
\newblock {\em Demography\/}~{\em 46\/}(1), 103--125.

\bibitem[\protect\citeauthoryear{Hanneke, Fu, Xing, et~al.}{Hanneke
  et~al.}{2010}]{hanneke2010discrete}
Hanneke, S., W.~Fu, E.~P. Xing, et~al. (2010).
\newblock Discrete temporal models of social networks.
\newblock {\em Electronic Journal of Statistics\/}~{\em 4}, 585--605.

\bibitem[\protect\citeauthoryear{Hoff}{Hoff}{2021}]{hoff2021additive}
Hoff, P. (2021).
\newblock Additive and multiplicative effects network models.
\newblock {\em Statistical Science\/}~{\em 36\/}(1), 34--50.

\bibitem[\protect\citeauthoryear{Hoff}{Hoff}{2005}]{hoff2005bilinear}
Hoff, P.~D. (2005).
\newblock Bilinear mixed-effects models for dyadic data.
\newblock {\em Journal of the American Statistical Association\/}~{\em
  100\/}(469), 286--295.

\bibitem[\protect\citeauthoryear{Hoff, Raftery, and Handcock}{Hoff
  et~al.}{2002}]{hoff2002latent}
Hoff, P.~D., A.~E. Raftery, and M.~S. Handcock (2002).
\newblock Latent space approaches to social network analysis.
\newblock {\em Journal of the American Statistical Association\/}~{\em
  97\/}(460), 1090--1098.

\bibitem[\protect\citeauthoryear{Hunter, Goodreau, and Handcock}{Hunter
  et~al.}{2008}]{hunter2008goodness}
Hunter, D.~R., S.~M. Goodreau, and M.~S. Handcock (2008).
\newblock Goodness of fit of social network models.
\newblock {\em Journal of the American Statistical Association\/}~{\em
  103\/}(481), 248--258.

\bibitem[\protect\citeauthoryear{Kim, Lee, Xue, Niu, et~al.}{Kim
  et~al.}{2018}]{kim2017review}
Kim, B., K.~H. Lee, L.~Xue, X.~Niu, et~al. (2018).
\newblock A review of dynamic network models with latent variables.
\newblock {\em Statistics Surveys\/}~{\em 12}, 105--135.

\bibitem[\protect\citeauthoryear{Kim, Niu, Hunter, and Cao}{Kim
  et~al.}{2023}]{bomin23}
Kim, B., X.~Niu, D.~Hunter, and X.~Cao (2023).
\newblock Supplement to ``a dynamic additive and multiplicative effects network
  model with application to the united nations voting behaviors''.

\bibitem[\protect\citeauthoryear{Krivitsky and Handcock}{Krivitsky and
  Handcock}{2014}]{stergm}
Krivitsky, P.~N. and M.~S. Handcock (2014).
\newblock A separable model for dynamic networks.
\newblock {\em Journal of the Royal Statistical Society. Series B, Statistical
  Methodology\/}~{\em 76\/}(1), 29--46.

\bibitem[\protect\citeauthoryear{Leibenstein}{Leibenstein}{1966}]{leibenstein1966shaping}
Leibenstein, H. (1966).
\newblock Shaping the world economy: Suggestions for an international economic
  policy.

\bibitem[\protect\citeauthoryear{Lusher, Koskinen, and Robins}{Lusher
  et~al.}{2013}]{lusher2013}
Lusher, D., J.~Koskinen, and G.~Robins (2013).
\newblock {\em Exponential random graph models for social networks: Theory,
  methods, and applications}.
\newblock Cambridge University Press.

\bibitem[\protect\citeauthoryear{Marshall, Jaggers, and Gurr}{Marshall
  et~al.}{2014}]{marshall2014polity}
Marshall, M.~G., K.~Jaggers, and T.~R. Gurr (2014).
\newblock Polity {IV} annual time-series, 1800--2013.
\newblock {\em Center for International Development and Conflict Management at
  the University of Maryland College Park\/}.

\bibitem[\protect\citeauthoryear{Olivella, Pratt, and Imai}{Olivella
  et~al.}{2022}]{dsbm22}
Olivella, S., T.~Pratt, and K.~Imai (2022).
\newblock Dynamic stochastic blockmodel regression for network data:
  Application to international militarized conflicts.
\newblock {\em Journal of the American Statistical Association\/}.

\bibitem[\protect\citeauthoryear{Rasmussen}{Rasmussen}{2003}]{rasmussen2004gaussian}
Rasmussen, C.~E. (2003).
\newblock {G}aussian processes in machine learning.
\newblock In {\em Summer School on Machine Learning}, pp.\  63--71. Springer.

\bibitem[\protect\citeauthoryear{Rodrigue, Comtois, and Slack}{Rodrigue
  et~al.}{2016}]{rodrigue2009geography}
Rodrigue, J.-P., C.~Comtois, and B.~Slack (2016).
\newblock {\em The geography of transport systems}.
\newblock Routledge.

\bibitem[\protect\citeauthoryear{Sewell and Chen}{Sewell and
  Chen}{2015}]{sewell2015latent}
Sewell, D.~K. and Y.~Chen (2015).
\newblock Latent space models for dynamic networks.
\newblock {\em Journal of the American Statistical Association\/}~{\em
  110\/}(512), 1646--1657.

\bibitem[\protect\citeauthoryear{Signorino and Ritter}{Signorino and
  Ritter}{1999}]{signorino1999tau}
Signorino, C.~S. and J.~M. Ritter (1999).
\newblock Tau-b or not tau-b: Measuring the similarity of foreign policy
  positions.
\newblock {\em International Studies Quarterly\/}~{\em 43\/}(1), 115--144.

\bibitem[\protect\citeauthoryear{Snijders, Van~de Bunt, and Steglich}{Snijders
  et~al.}{2010}]{snijders2010introduction}
Snijders, T.~A., G.~G. Van~de Bunt, and C.~E. Steglich (2010).
\newblock Introduction to stochastic actor-based models for network dynamics.
\newblock {\em Social Networks\/}~{\em 32\/}(1), 44--60.

\bibitem[\protect\citeauthoryear{Snijders}{Snijders}{2001}]{snijders01}
Snijders, T. A.~B. (2001).
\newblock The statistical evaluation of social network dynamics.
\newblock {\em Sociological Methodology\/}~{\em 31\/}(1), 361--395.

\bibitem[\protect\citeauthoryear{Voeten}{Voeten}{2013}]{voeten13}
Voeten, E. (2013).
\newblock Data and analyses of voting in the {UN} general assembly.
\newblock In B.~Reinalda (Ed.), {\em Routledge Handbook of International
  Organization}. Routledge.
\newblock Available at SSRN: http://ssrn.com/abstract=2111149.

\bibitem[\protect\citeauthoryear{Ward and Hoff}{Ward and
  Hoff}{2007}]{ward2007persistent}
Ward, M.~D. and P.~D. Hoff (2007).
\newblock Persistent patterns of international commerce.
\newblock {\em Journal of Peace Research\/}~{\em 44\/}(2), 157--175.

\bibitem[\protect\citeauthoryear{Zellner}{Zellner}{1986}]{zellner86}
Zellner, A. (1986).
\newblock On assessing prior distributions and bayesian regression analysis
  with g-prior distributions.
\newblock In P.~K. Goel and A.~Zellner (Eds.), {\em Bayesian inference and
  decision techniques: Essays in Honor of Bruno de Finetti (Studies in Bayesian
  Econometrics and Statistics, Vol 6)}, pp.\  233--243. Amsterdam:
  North-Holland.

\end{thebibliography}

\end{document}


\begin{center}
    {\LARGE\bf Supplementary Materials to: ``A Dynamic Additive and Multiplicative Effects Network Model
	with Application to the United Nations Voting Behaviors ''}
\end{center}

\begin{appendix}

\section{Derivations of Full Conditional Distributions}\label{subsec:derivation}
The noise error variance $\sigma_e^2$ has a conditional inverse-gamma distribution with parameters as derived below:
\begin{equation*}
\begin{aligned}
&\Pr(\sigma_e^2|\mathbf{Y}, k_\sigma, \theta_\sigma) \propto\Pr(\mathbf{Y}|\mathbf{X}, \boldsymbol{\beta}, \boldsymbol{a}, \boldsymbol{d}, \boldsymbol{u},\sigma_e^2)\times \Pr(\sigma_e^2|k_\sigma, \theta_\sigma)\\
&\propto\prod\limits_{t=1}^T\prod\limits_{i> j}(\sigma_e^2)^{-\frac{1}{2}}\mbox{exp}\Big\{-\frac{1}{2\sigma_e^2}||y^t_{ij}-\big(\sum\limits_{p=1}^P \beta^t_{p}X^t_{ijp}+a^t_{i}+a^t_{j}+{\boldsymbol{u}^t_{i}}^\prime \mathbf{D}^t\boldsymbol{u}^t_{j}\big)||^2\Big\}\times (\sigma_e^2)^{-k_\sigma-1}\mbox{exp}\Big\{\frac{1}{\sigma_e^2}\theta_\sigma\Big\}\\
&=(\sigma_e^2)^{-\frac{T}{2}\cdot\frac{N(N-1)}{2}-k_\sigma-1}\times\mbox{exp}\Big\{-\frac{1}{\sigma_e^2}\Big(\frac{1}{2}\sum\limits_{t=1}^T\sum\limits_{i> j}||y^t_{ij}-\big(\sum\limits_{p=1}^P \beta^t_{p}X^t_{ijp}+a^t_{i}+a^t_{j}+{\boldsymbol{u}^t_{i}}^\prime \mathbf{D}^t\boldsymbol{u}^t_{j}\big)||^2+\theta_\sigma\Big)\Big\}\\
&\sim \mathcal{IG}\left(\frac{T\cdot N(N-1)}{4}+k_\sigma,\quad \frac{1}{2}\sum\limits_{t=1}^T\sum\limits_{i> j}(E^t_{ij})^2+\theta_\sigma\right)	
\end{aligned}
\end{equation*} 
The fixed-effect coefficient $\boldsymbol{\beta}_p$, for $1\le p\le P$, has a conditional multivariate normal distribution with parameters as derived below:
\begin{equation*}
\begin{aligned}
			&\Pr(\boldsymbol{\beta}_p|\mathbf{Y}, \mathbf{X}, \kappa^\beta_p, \tau^\beta_p) \propto \Pr(\mathbf{Y}|\mathbf{X}, \boldsymbol{\beta}_p, \boldsymbol{\beta}_{[-p]}, \boldsymbol{a}, \boldsymbol{d}, \boldsymbol{u},\sigma_e^2) \times \Pr(\boldsymbol{\beta}_p|\kappa^\beta_p, \tau^\beta_p) \\
			&\propto\prod\limits_{i>j}\mbox{exp}\Big\{-\frac{1}{2\sigma_e^2}||\mathbf{E}_{ij[-p]}-\mathbf{X}_{ijp}\boldsymbol{\beta}_p||^2\Big\}\times \mbox{exp}\Big\{-\frac{1}{2}\big(\boldsymbol{\beta}'_p(\Sigma^\beta_p)^{-1}\boldsymbol{\beta}_p\big)\Big\} \\ &\quad\quad\mbox{ where } \mathbf{E}_{ij[-p]} = \{E^t_{ij[-p]}\}_{t=1}^T \mbox{ (with }E^{t}_{ij[-p]}=E^t_{ij}+\beta^t_{p}X^{t}_{ijp})\mbox{ and } \mathbf{X}_{ijp} = \{X^t_{ijp}\}_{t=1}^T\\
			&\propto\mbox{exp}\Big\{-\frac{1}{2\sigma_e^2}\Big(\sum\limits_{i>j}-2(\mathbf{E}_{ij[-p]}\mathbf{X}_{ijp})'\boldsymbol{\beta}_p+\boldsymbol{\beta}'_p\big(\mbox{diag}(\sum\limits_{i>j}{\mathbf{X}^2_{ijp}})\big)\boldsymbol{\beta}_p\Big)\Big\}\times \mbox{exp}\Big\{-\frac{1}{2}\big(\boldsymbol{\beta}'_p(\Sigma^\beta_p)^{-1}\boldsymbol{\beta}_p\big)\Big\}\\
			&\propto\mbox{exp}\Big\{-\frac{1}{2}\Big(\boldsymbol{\beta}'_p\big((\Sigma^\beta_p)^{-1}+\frac{\mbox{diag}(\sum_{i>j}{\mathbf{X}^2_{ijp}})}{\sigma_e^2}\big)\boldsymbol{\beta}_p-\frac{2}{\sigma_e^2}\big(\sum_{i>j}(\mathbf{E}_{ij[-p]}\mathbf{X}_{ijp})'\boldsymbol{\beta}_p\big)\Big)\Big\}\\
			& \sim \mathcal{N}_T\big(\tilde{\mu}_{\beta_p}, \tilde{\Sigma}_{\beta_p} \big),\\
			&\mbox{ where } \tilde{\Sigma}_{\beta_p} = \left[(\Sigma^\beta_p)^{-1}+\frac{\mbox{diag}\big(\{\sum_{i>j}{X^{t2}_{ijp}}\}_{t=1}^{T}\big)}{\sigma_e^2}\right]^{-1} \mbox{ and } \tilde{\mu}_{\beta_p} =  \left[\frac{\{\sum_{i>j}(E^{t}_{ij[-p]}X^t_{ijp})\}_{t=1}^{T}}{\sigma_e^2}\right]\tilde{\Sigma}_{\beta_p}.
\end{aligned}
\end{equation*} 
The $T$-dimensional additive random effect $\boldsymbol{a}_i$, for $1\le i\le N$, has a conditional multivariate normal distribution with parameters as derived below:
\begin{equation*}
\begin{aligned}
		&\Pr(\boldsymbol{a}_{i}|\mathbf{Y}, \kappa^a, \tau^a) \propto \prod\limits_{\substack{ j\neq i}}\Pr(\mathbf{Y}|\mathbf{X},\boldsymbol{\beta}, \boldsymbol{a}_{i}, \boldsymbol{a}_{[-i]}, \boldsymbol{d}, \boldsymbol{u},\sigma_e^2) \times \Pr(\boldsymbol{a}_{i}|\kappa^a, \tau^a) \\
		&\propto\prod\limits_{\substack{j\neq i}}\mbox{exp}\Big\{-\frac{1}{2\sigma_e^2}||\mathbf{E}_{ij[-i]}-\boldsymbol{a}_{i}||^2\Big\}\times \mbox{exp}\Big\{-\frac{1}{2}\big(\boldsymbol{a}'_{i}{(\Sigma^a)}^{-1}\boldsymbol{a}_{i}\big)\Big\}\\
		& \quad\quad\mbox{where } \mathbf{E}_{ij[-i]}=\{E^t_{ij[-i]}\}_{t=1}^T \mbox{ with } E^t_{ij[-i]}=E^t_{ij}+\boldsymbol{a}^t_{i}\\
		&\propto\mbox{exp}\Big\{-\frac{1}{2\sigma_e^2}\Big(\sum\limits_{\substack{j\neq i}}-2(\mathbf{E}_{ij[-i]})'\boldsymbol{a}_{i}+\boldsymbol{a}'_{i}\big(\sum\limits_{\substack{ j\neq i}}I_T\big)\boldsymbol{a}_{i}\Big)\Big\}\times \mbox{exp}\Big\{-\frac{1}{2}\big(\boldsymbol{a}'_{i}{(\Sigma^a)}^{-1}\boldsymbol{\theta}_{i}\big)\Big\}\\
		&\propto\mbox{exp}\Big\{-\frac{1}{2}\Big(\boldsymbol{a}'_{i}\big({(\Sigma^a)}^{-1}+\frac{(N-1)I_T}{\sigma_e^2}\big)\boldsymbol{a}_{i}-\frac{2}{\sigma_e^2}\big(\sum_{ j\neq i}(\mathbf{E}_{ij[-i]})'\boldsymbol{a}_{i}\big)\Big)\Big\}\\
		& \sim \mathcal{N}_T\big(\tilde{\mu}_{a_i}, \tilde{\Sigma}_{a_i} \big),\\
		& \mbox{ where } \tilde{\Sigma}_{a_i} = \left[ (\Sigma^a)^{-1}+\frac{(N-1)I_T}{\sigma_e^2}\right]^{-1} \mbox{ and }
		\tilde{\mu}_{a_i} = \left[ \frac{\{\sum_{j\neq i}E^{t}_{ij[-i]}\}_{t=1}^{T}}{\sigma_e^2}\right] \tilde{\Sigma}_{a_i}.
\end{aligned}
\end{equation*} 
The $T$-dimensional latent vector of the multiplicative random effect $\boldsymbol{u}_{ir}$, for $1\le i \le N$ and $1\le r\le R$, has a conditional multivariate normal distribution with parameters as derived below:
\begin{equation*}
	\begin{aligned}
	&\Pr(\boldsymbol{u}_{ir}|\mathbf{Y}, \kappa_r^u, \tau_r^u) \propto \Pr(\mathbf{Y}|\mathbf{X}, \boldsymbol{\beta}, \boldsymbol{a}, \boldsymbol{d}, \boldsymbol{u}_{ir},\boldsymbol{u}_{[-ir]}, \sigma_e^2) \times \Pr(\boldsymbol{u}_{ir}| \kappa_r^u, \tau_r^u) \\
	&\propto\prod\limits_{i=1}^N\prod\limits_{j\neq i}\mbox{exp}\Big\{-\frac{1}{2\sigma_e^2}||\mathbf{E}_{ij[-r]}-\boldsymbol{u}_{ir}^\prime \boldsymbol{d}_r\boldsymbol{u}_{jr}||^2\Big\}\times \mbox{exp}\Big\{-\frac{1}{2}\big({\boldsymbol{u}_{ir}}^\prime(\Sigma_r^u)^{-1}\boldsymbol{u}_{ir}\big)\Big\}\\
	&\quad\quad\mbox{ where } \mathbf{E}_{ij[-r]} = \{E^t_{ij[-r]}\}_{t=1}^T \mbox{ (with }E^{t}_{ij[-r]}=E^t_{ij}+{u^t_{ir}}^\prime d_r^t u^t_{ir})\mbox{ and } \boldsymbol{u}_{ir}^\prime \boldsymbol{d}_r\boldsymbol{u}_{jr} = \{{u^t_{ir}}^\prime d_r^t u^t_{ir}\}_{t=1}^T\\
	&	\propto\mbox{exp}\Big\{-\frac{1}{4\sigma_e^2}\Big(\sum\limits_{j \neq i }-2(\mathbf{E}_{ij[-r]}\boldsymbol{d}_{r}\boldsymbol{u}_{jr})'\boldsymbol{u}_{ir}+\boldsymbol{u}'_{ir}\big(\mbox{diag}(\boldsymbol{d}_{r}^2\sum\limits_{j \neq i}{\boldsymbol{u}^2_{jr}})\big)\boldsymbol{u}_{ir}\Big)\Big\}\times \mbox{exp}\Big\{-\frac{1}{2}\big({\boldsymbol{u}_{ir}}^\prime(\Sigma_r^u)^{-1}\boldsymbol{u}_{ir}\big)\Big\}\\
	&\propto\mbox{exp}\Big\{-\frac{1}{2}\Big(\boldsymbol{u}'_{ir}\big((\Sigma_r^u)^{-1}+\frac{1}{2}\times\frac{\mbox{diag}(\boldsymbol{d}_{r}^2\sum_{j \neq i}{\boldsymbol{u}^2_{jr}})}{\sigma_e^2}\big)\boldsymbol{u}_{ir}-\frac{2}{\sigma_e^2}\big(\frac{1}{2}\sum_{j \neq i}(\mathbf{E}_{ij[-r]}\boldsymbol{d}_{r}\boldsymbol{u}_{jr})'\boldsymbol{u}_{ir}\big)\Big)\Big\}\\
	& \sim \mathcal{N}_T\big(\tilde{\mu}_{u_{ir}}, \tilde{\Sigma}_{u_{ir}} \big),\\
	& \mbox{ where }\tilde{\Sigma}_{u_{ir}} = \left[ (\Sigma_r^u)^{-1}+\frac{\mbox{diag}\big(\{({{d}^t_{r}})^2\sum_{j \neq i}{({{u}^t_{jr}}})^2\}_{t=1}^{T}\big)}{2\sigma_e^2}\right]^{-1} \mbox{ and } \tilde{\mu}_{u_{ir}} =  \left[ \frac{\{\sum_{j \neq i}(E^{t}_{ij[-r]}d^t_{r}u^t_{jr})\}_{t=1}^{T}}{2\sigma_e^2}\right] \tilde{\Sigma}_{u_{ir}}.
	\end{aligned}
\end{equation*} 
The $T$-dimensional vector of the $r$th eigenvalue of the multiplicative random effect $\boldsymbol{d}_r$, for $1\le r\le R$, 
has a conditional multivariate normal distribution with parameters as derived below:
\begin{equation*}
\begin{aligned}
		&\Pr(\boldsymbol{d}_r|\mathbf{Y}) \propto \Pr(\mathbf{Y}|\mathbf{X}, \boldsymbol{\beta}, \boldsymbol{a}, \boldsymbol{d}_r, \boldsymbol{d}_{[-r]}, \boldsymbol{u},\sigma_e^2) \times \Pr(\boldsymbol{d}_r) \\
		&\propto\prod\limits_{i>j}\mbox{exp}\Big\{-\frac{1}{2\sigma_e^2}||\mathbf{E}_{ij[-r]}-\boldsymbol{u}_{ir}^\prime \boldsymbol{d}_r\boldsymbol{u}_{jr}||^2\Big\}\times \mbox{exp}\Big\{-\frac{1}{2}\big(\boldsymbol{d}'_r\boldsymbol{d}_r\big)\Big\}\\ &\quad\quad\mbox{ where } \mathbf{E}_{ij[-r]} = \{E^t_{ij[-r]}\}_{t=1}^T \mbox{ (with }E^{t}_{ij[-r]}=E^t_{ij}+{u^t_{ir}}^\prime d_r^t u^t_{ir})\mbox{ and } \boldsymbol{u}_{ir}^\prime \boldsymbol{d}_r\boldsymbol{u}_{jr} = \{{u^t_{ir}}^\prime d_r^t u^t_{ir}\}_{t=1}^T\\
		&\propto\mbox{exp}\Big\{-\frac{1}{2\sigma_e^2}\Big(\sum\limits_{i>j}-2(\mathbf{E}_{ij[-r]}\boldsymbol{u}_{ir}\boldsymbol{u}_{jr})'\boldsymbol{d}_r+\boldsymbol{d}'_r\big(\mbox{diag}(\sum\limits_{i>j}({\boldsymbol{u}_{ir}\boldsymbol{u}_{jr}})^2)\big)\boldsymbol{d}_r\Big)\Big\}\times \mbox{exp}\Big\{-\frac{1}{2}\big(\boldsymbol{d}'_r\boldsymbol{d}_r\big)\Big\}\\
		&\propto\mbox{exp}\Big\{-\frac{1}{2}\Big(\boldsymbol{d}'_r\big(I_T+\frac{\mbox{diag}(\sum_{i>j}({\boldsymbol{u}_{ir}\boldsymbol{u}_{jr}})^2)}{\sigma_e^2}\big)\boldsymbol{d}_r-\frac{2}{\sigma_e^2}\big(\sum_{i>j}(\mathbf{E}_{ij[-r]}\boldsymbol{u}_{ir}\boldsymbol{u}_{jr})'\boldsymbol{d}_r\big)\Big)\Big\}\\
		& \sim \mathcal{N}_T\big(\tilde{\mu}_{d_r}, \tilde{\Sigma}_{d_r} \big),\\
		& \mbox{ where }\tilde{\Sigma}_{d_r} = \left[ I_T+\frac{\mbox{diag}\big(\{\sum_{i>j}({u^t_{ir}u^t_{jr}})^2\}_{t=1}^{T}\big)}{\sigma_e^2}\right]^{-1} \mbox{ and } \tilde{\mu}_{d_r} =  \left[ \frac{\{\sum_{i>j}(E^{t}_{ij[-r]}u^t_{ir}u^t_{jr})\}_{t=1}^{T}}{\sigma_e^2}\right] \tilde{\Sigma}_{d_r}.
		\end{aligned}
\end{equation*} 

\section{Metropolis-Hastings Algorithm for GP Parameters}\label{subsec:MH}
For the Gaussian process variance parameter $\tau$ and length-scale parameter $\kappa$, we use the Metropolis-Hastings algorithm with a proposal 
density $Q$ being the multivariate Gaussian distribution with a diagonal covariance matrix---i.e., diag$ (\sigma^2_{Q1}, \sigma^2_{Q2})$. Given the proposal variance $\sigma^2_Q = (\sigma^2_{Q1}, \sigma^2_{Q2})$, we sample the new values $\tau^\prime$ and $\kappa^\prime$ from
\begin{equation*}
(p^\prime, q^\prime) \sim  \mathcal{N}_2((p, q), \mbox{diag}(\sigma^2_Q) ),
\end{equation*}
where $p= \log(\tau)$ and $q=\log(\kappa)$. We use change of variables and sample from a bivariate normal with mean $(p, q) = \log(\tau, \kappa)$ since both $\tau$ and $\kappa$ have to be positive. 
Under the symmetric proposal distribution as above, we accept the new proposed value $(\tau^\prime, \kappa^\prime) = (\mbox{exp}(p^\prime),\mbox{exp}( q^\prime))$ with probability equal to
\begin{equation*}
\min\left\{1,  \frac{\Pr(p^{\eta\prime}, q^{\eta\prime}|\,\boldsymbol{\eta}, k_\eta, \theta_\eta,\gamma)}{\Pr(p^{\eta}, q^{\eta}|\,\boldsymbol{\eta}, k_\eta, \theta_\eta,\gamma)} \right\},
\end{equation*}
where $\boldsymbol{\eta}=(\eta_1,\ldots,\eta_T)$ can be any of $\boldsymbol{\beta}$, $\boldsymbol{a}$, or $\boldsymbol{u}$, depending on the situation. After taking account of the Jacobian, $\Pr_{p,q}(p^{\eta}, q^{\eta}|\,\boldsymbol{\eta}, k_\eta, \theta_\eta,\gamma) =\Pr_{\tau,\kappa}((\exp(p^{\eta}), (\exp(q^{\eta}))|\,\boldsymbol{\eta}, k_\eta, \theta_\eta,\gamma) \times \exp(p^{\eta}) \times \exp(q^{\eta}).$

Below are the acceptance ratios for all pairs of variables. In each case, $\tau$ has an $\mathcal{IG}(k_\eta,\theta_\eta)$ prior and $\kappa$ has an independent 
$\mbox{half-Cauchy}(\gamma_\eta)$ prior.
\begin{itemize}
	\item [1.] $(\tau^\beta_p, \kappa^\beta_p), \mbox{ for } p=1,\ldots,P$:
	\begin{equation*}
	\begin{aligned}
	\frac{\Pr(\tau^{\beta\prime}_p, \kappa^{\beta\prime}_p|\,\boldsymbol{\beta}_p, k_\beta,\theta_\beta,\gamma_\beta)\times\tau^{\beta\prime}_p \kappa^{\beta\prime}_p}{\Pr(\tau^\beta_p, \kappa^\beta_p|\,\boldsymbol{\beta}_p, k_\beta, \theta_\beta,\gamma_\beta)\times \tau^\beta_p \kappa^\beta_p}=&\frac{\Pr(\tau^{\beta\prime}_p, \kappa^{\beta\prime}_p, \boldsymbol{\beta}_p|\, k_\beta, \theta_\beta,\gamma_\beta)\times\tau^{\beta\prime}_p \kappa^{\beta\prime}_p}{\Pr(\tau^\beta_p, \kappa^\beta_p,\boldsymbol{\beta}_p|\,k_\beta, \theta_\beta,\gamma_\beta)\times \tau^\beta_p \kappa^\beta_p}\\=&\frac{\Pr(\tau^{\beta\prime}_p|\,k_\beta,\theta_\beta)\Pr(\kappa^{\beta\prime}_p|\,\gamma_\beta)\Pr( \boldsymbol{\beta}_p|\,\tau^{\beta\prime}_p, \kappa^{\beta\prime}_p) \times\tau^{\beta\prime}_p \kappa^{\beta\prime}_p}{\Pr(\tau^{\beta}_p|\,k_\beta, \theta_\beta)\Pr(\kappa^{\beta}_p|\,\gamma_\beta)\Pr( \boldsymbol{\beta}_p|\,\tau^{\beta}_p, \kappa^{\beta}_p)\times \tau^\beta_p \kappa^\beta_p},
	\end{aligned}
	\end{equation*}
	where $\boldsymbol{\beta}_{p}\sim \mathcal{N}_T(\boldsymbol{0}, \Sigma^\beta_p)$ with $\Sigma^\beta_p = \tau_p^\beta f(\kappa^\beta_{p})$.
	\item [2.] $(\tau^a, \kappa^a)$
	\begin{equation*}
	\begin{aligned}
	\frac{\Pr(\tau^{a\prime}, \kappa^{a\prime}|\,\boldsymbol{a}, k_a, \theta_a,\gamma_a)\times\tau^{a\prime} \kappa^{a\prime}}{\Pr(\tau^a, \kappa^a|\,\boldsymbol{a}, k_a, \theta_a,\gamma_a)\times\tau^a \kappa^a}=&\frac{\Pr(\tau^{a\prime}, \kappa^{a\prime}, \boldsymbol{a}|\, k_a,\theta_a,\gamma_a)\times\tau^{a\prime} \kappa^{a\prime}}{\Pr(\tau^a, \kappa^a,\boldsymbol{a}|\,k_a, \theta_a,\gamma_a)\times\tau^a \kappa^a}\\=&\frac{\Pr(\tau^{a\prime}|\,k_a, \theta_a)\Pr(\kappa^{a\prime}|\,\gamma_a)\prod_{i=1}^N \Pr(\boldsymbol{a}_i|\,\tau^{a\prime}, \kappa^{a\prime}) \times\tau^{a\prime} \kappa^{a\prime}}{\Pr(\tau^{a}|\,k_a, \theta_a)\Pr(\kappa^{a}|\,\gamma_a)\prod_{i=1}^N\Pr( \boldsymbol{a}_i|\,\tau^{a}, \kappa^{a})\times\tau^a \kappa^a},\\
	\end{aligned}
	\end{equation*}
	where $	\boldsymbol{a}_{i}\sim \mathcal{N}_T(\boldsymbol{0},\Sigma^a)$ with $\Sigma^a = \tau^a f(\kappa^a)$.
	\item [3.] $(\tau_{r}^u, \kappa_{r}^u),$ for $r=1,\ldots,R$:
	\begin{equation*}
	\begin{aligned}
	\frac{\Pr(\tau_r^{u\prime}, \kappa_r^{u\prime}|\,\boldsymbol{u}_r, k_u, \theta_u,\gamma_u)\times \tau_r^{u\prime} \kappa_r^{u\prime}}{\Pr(\tau_r^u, \kappa^u|\,\boldsymbol{u}_r, k_u, \theta_u,\gamma_u)\times \tau_r^u\kappa^u}=&\frac{\Pr(\tau_r^{u\prime}, \kappa_r^{u\prime}, \boldsymbol{u}_r| \,k_u, \theta_u,\gamma_u)\times \tau_r^{u\prime} \kappa_r^{u\prime}}{\Pr( \tau_r^u, \kappa_r^u, \boldsymbol{u}_r|\, k_u, \theta_u,\gamma_u)\times \tau_r^u\kappa^u}\\=&\frac{\Pr(\tau_r^{u\prime}|\, k_u, \theta_u)\Pr(\kappa_r^{u\prime}|\,\gamma_u)\prod_{i=1}^N\Pr(\boldsymbol{u}_{ir}|\,\tau_r^{u\prime}, \kappa_r^{u\prime}) \times \tau_r^{u\prime} \kappa_r^{u\prime}}{\Pr(\tau_r^u|\, k_u,\theta_u)\Pr(\kappa_r^{u}|\,\gamma_u)\prod_{i=1}^N\Pr( \boldsymbol{u}_{ir}|\, \tau_r^{u},\kappa_r^{u})\times \tau_r^u\kappa^u},
	\end{aligned}
	\end{equation*}
	where $\boldsymbol{u}_{ir}\sim \mathcal{N}_T(\boldsymbol{0}, \Sigma^u_r)$ with $\Sigma^u_r = \tau_r^u f(\kappa^u_{r})$.
\end{itemize}

\section{List of Countries in Voting Network}\label{appendix: countrylist}
\begin{table}[H]
\centering
\scalebox{0.63}{
\begin{tabular}{c|c||c|c}
	\hline
	Abbreviation&Country or area name &	Abbreviation&Country or area name\\
	\hline
			AFG* & Afghanistan & 	KUW & Kuwait\\
				ALB & Albania & LEB* & Lebanon\\
			ALG & Algeria & LIB &Libya \\
		ANG & Angola 	& MAA&  Mauritania\\
		ARG& Argentina	&MEX &Mexico \\
		AUL* & Australia&MLI&Mali\\
		BAH & Bahrain	&MOR & Morocco\\
	BEN & Benin 	&MZM & Mozambique\\
			BFO & Burkina Faso 	&NEW & New Zealand \\
	BNG&  Bangladesh	& NIC& Nicaragua\\
			BOL& Bolivia	& NIG &      Nigeria\\
		BRA &Brazil&NIR & Niger\\
				BUI & Burundi &NOR &Norway \\
		BUL & Bulgaria	&NTH &  Netherlands\\
		CAN & Canada	& OMA & Oman\\
		CAO & Cameroon	&PAK* & Pakistan\\
		CEN & Central African Republic	&PAN & Panama\\
			CHL & Chile &PAR & Paraguay\\
			CHN* & China	& PER & Peru \\
		COL & Colombia	&PHI & Philippines\\
			CON & Congo		&POL & Poland\\
			COS & Costa Rica&POR & Portugal\\
			DEN& Denmark	 & PRK* & North Korea\\
			DOM & Dominican Republic	&QAT & Qatar\\
		ECU &Ecuador 	&ROK* & South Korea\\
			EGY* & Egypt	&RUS* & Russia\\
				FIN & Finland	&RWA & Rwanda\\
			FRN* & France&SAL & El Salvador\\
			GAB & Gabon 	&SAU & Saudi Arabia\\
			GAM & Gambia	&SEN &Senegal \\
				GMY* & Germany& SIE&Sierra Leone \\
			 	GHA & Ghana	&SPN & Spain\\
			GRC & Greece&SUD* &Sudan \\
				GUA & Guatemala &SUR&Suriname\\
				GUI &Guinea&SYR* & Syrian Arab Republic\\
		GUY & Guyana	&TAZ & Tanzania\\
				HAI & Haiti  &TOG & Togo\\
					HON &Honduras&TRI & Trinidad and Tobago\\
			HUN & Hungary 	&TUN & Tunisia\\
				IND* & India&TUR*& Turkey\\
			INS & Indonesia		&UAE & United Arab Emirates\\
	IRN* & Iran (Islamic Republic of)		& UGA & Uganda\\
			IRQ* & Iraq &UKG* & United Kingdom\\
				ISR* & Israel&URU & Uruguay\\
			ITA & Italy		&USA* & United States of America\\
		JAM& Jambia	&VEN & Venezuela\\
			JOR & Jordan	&ZAM & Zambia\\
			JPN* & Japan	&ZIM & Zimbabwe\\		
				KEN & Kenya&&\\
				\hline
\end{tabular}}
	\label{table:importantvotes}
	\caption {Full list of the 97 countries, where the 21 most active countries during the ten year period 2004--2014 according to \citet{hoff2015multilinear} are marked with $*$.}
\end{table}

\section{Summary Statistics}\label{subsec:EDAsummary}
\begin{table}[H]
	\centering
	\begin{tabular}{ |c|c|c|c|c|c|c|c|c|c|} 
		\hline
		{Year}	& 1983 & 1984& 1985& 1986 & 1987& 1988& 1989&1990&1991\\ \hline
		Joint votes & 126.709&128.302& 134.563& 139.387& 133.998& 123.612& 104.472&  78.943 & 61.753\\\hline
		Agreement & 84.7& 85.2&84.3& 85.4& 87.7& 87.2& 88.5& 87.8&86.4\\\hline\hline 
		{Year}	& 1992&1993& 1994 & 1995& 1996& 1997 & 1998& 1999& 2000\\ \hline
		Joint votes  & 58.263&  52.001&  56.187& 62.019&  61.477&  58.104&  50.969&55.071 & 52.731\\\hline
		Agreement &84.2&83.5& 84.7& 83.5& 84.4& 83.1& 85.1& 83.7& 83.9\\\hline\hline	
		{Year}	&2001&2002&2003&2004&2005& 2006& 2007& 2008&  2009\\ \hline
		Joint votes &  50.362& 58.282&  63.418&  61.860& 59.651& 73.235&64.857&62.353&  57.893\\\hline
		Agreement &81.5& 82.3& 83.0& 81.4& 83.5& 83.1& 81.9& 82.9& 80.4\\
		\hline\hline	
		Year & 2010& 2011&2012&2013&2014&&&&\\\hline
		Joint votes& 57.336&  54.404& 60.216& 56.016&  66.197&&&&\\\hline
		Agreement&  82.2& 79.8& 82.3& 80.2& 81.4&&&&\\\hline
	\end{tabular}
	\caption {Summary of the United Nations voting data for non-important votes: Average number of common votes (\textit{upper}) and averge voting similarity index (\textit{lower}) per year.}
	\label{table:unimportant}
\end{table}
\begin{table}[H]
	\centering
	\begin{tabular}{ |c|c|c|c|c|c|c|c|c|c|} 
		\hline
		{Year}	& 1983 & 1984& 1985& 1986 & 1987& 1988& 1989&1990&1991\\ \hline
		Joint votes & 7.384&7.441&8.129&9.201&8.283&5.121&12.605&7.361&8.559\\\hline
		Agreement & 69.6& 72.4& 72.5 & 74.8& 72.5& 77.3 & 79.8& 83.2& 83.6\\\hline\hline 
		{Year}	& 1992&1993& 1994 & 1995& 1996& 1997 & 1998& 1999& 2000\\ \hline
		Joint votes  &13.458&11.013&13.447&24.932&9.655&10.026&8.427&10.776&8.970\\\hline
		Agreement & 80.6& 77.3& 78.5& 82.7& 76.5& 80.5& 77.6& 82.9 & 76.8\\\hline\hline
		{Year}	&2001&2002&2003&2004&2005& 2006& 2007& 2008&  2009\\ \hline
		Joint votes &9.191&12.826&12.142&8.432&8,855&10.956&10.681&10.845&10.946\\\hline
		Agreement & 72.9& 81.6& 75.5& 83.5& 78.3& 73.3& 73.9& 70.0& 75.0\\		\hline\hline
		Year & 2010& 2011&2012&2013&2014&&&&\\\hline
		Joint votes&12.289&9.067&7.575&10.171&12.048&&&&\\\hline
		Agreement&  74.4& 75.1& 73.3& 75.4& 84.7&&&&\\\hline
	\end{tabular}
	\caption {Summary of the United Nations voting data for important votes: Average number of common votes (\textit{upper}) and averge voting similarity index (\textit{lower}) per year.}
	\label{table:EDA}
\end{table}
\begin{table}[H]
	\centering
	\begin{tabular}{ |c|ccccc|} 
		\hline
		correlation & 	log(distance) & polity &   alliance &  Trade/GDP  &  language\\
		\hline
		log(distance)& 1.000& 0.136&  -0.508&-0.355&-0.286\\
		polity&0.136&1.000& -0.264&-0.095&-0.142\\
		alliance& -0.508&-0.264&1.000&0.275& 0.417\\
		Trade/GDP&  -0.355& -0.095& 0.275& 1.000& 0.100\\
		language& -0.286&-0.142& 0.417& 0.100& 1.000\\
		\hline
	\end{tabular}
	\caption {Pearson correlation coefficients between the observed dyadic covariates.  Each
	variable is a ${97\choose 2} \times 32$-dimensional vector resulting from vectorizing
	across all years and all off-diagonal entries of the covariate matrices.  Two of the covariates, 
	$\log(\mbox{distance})$ and language, do not vary over time.}
	\label{table:correlation}
\end{table}

\section{Posterior Predictive Statistics for the United States (USA)}\label{appendix:USA}
\begin{figure}[H]
	\begin{center}
		\includegraphics[width=1\textwidth]{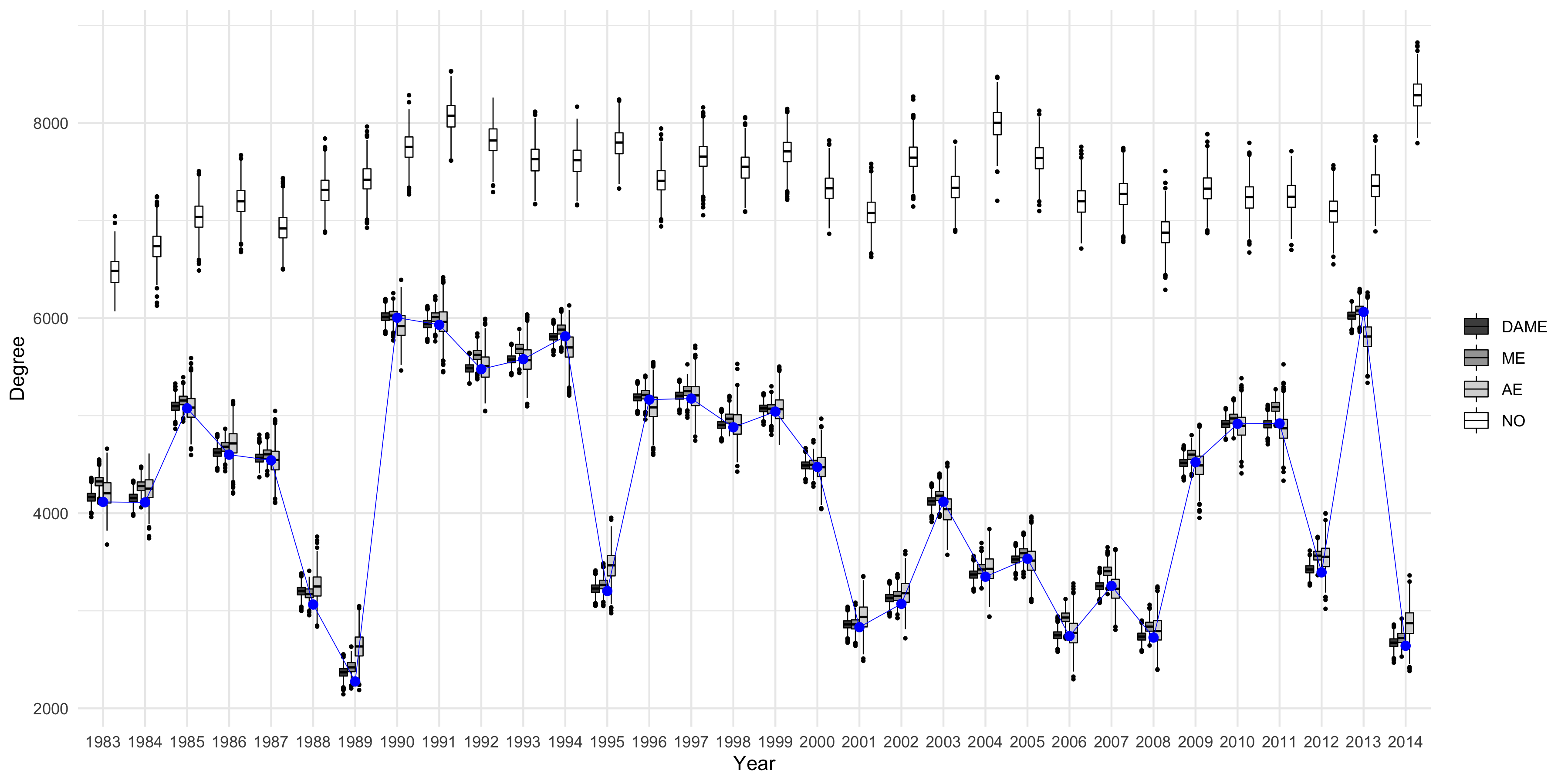}
	\end{center}
		\caption {Boxplots of 1,000 posterior predictive degree statistics for the 
United States (USA) for four models: The full DAME with both additive and multiplicative effects, the multiplicative effects only model ME, the additive effects
only model AE, and the model NO with neither. The dots show
the observed statistics.}
\label{figure:modelvalidation2}
		\label{figure:residual2}
\end{figure}

\section{Residual Diagnostics}\label{appendix: residual}
\begin{figure}[H]
	\begin{center}
		\includegraphics[width=0.8\textwidth]{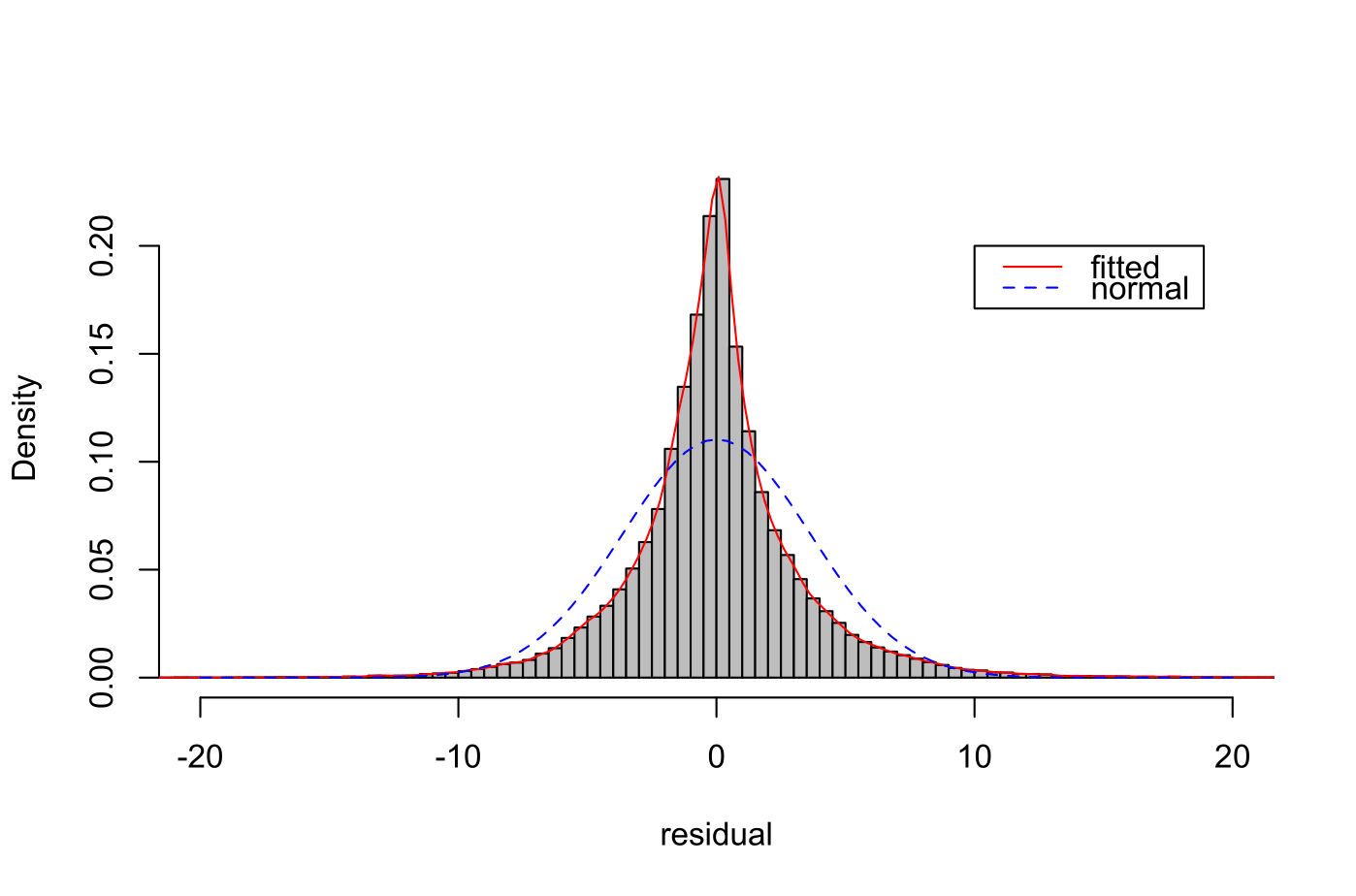}
	\end{center}
		\caption {Histogram of residuals from the DAME model (aggregating $\epsilon_{ij}^t$ for $1\le j < i\le 97$ and $1\le t\le 32$).}
		\label{figure:residual}
\end{figure}
As errors above seem closer to Laplacian than Normal distribution, we have tried an additional simulation based on Laplacian errors (while keeping other settings exactly the same as the simulation in Section 3.1 with medium correlation ($\kappa=2$), to examine the robustness of the DAME model to such error distribution misspecification. We find that if the errors are generated from Laplacian, our model performance is very similar to the normal error case. Therefore, our model seems to be robust when the errors are from Laplacian.  
\begin{figure}[H]
	\begin{center}
		\includegraphics[width=0.62\textwidth]{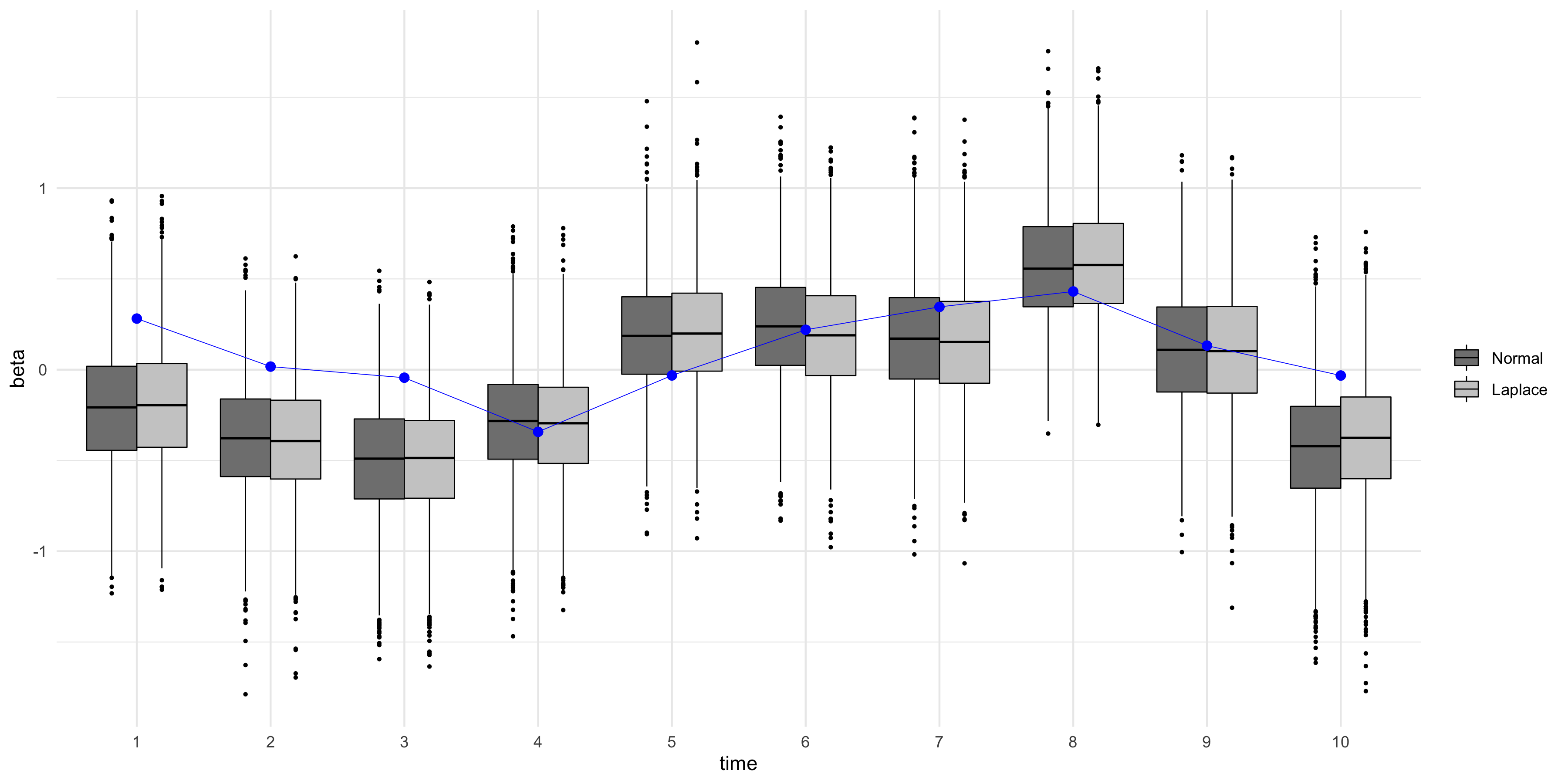}	
			\includegraphics[width=0.62\textwidth]{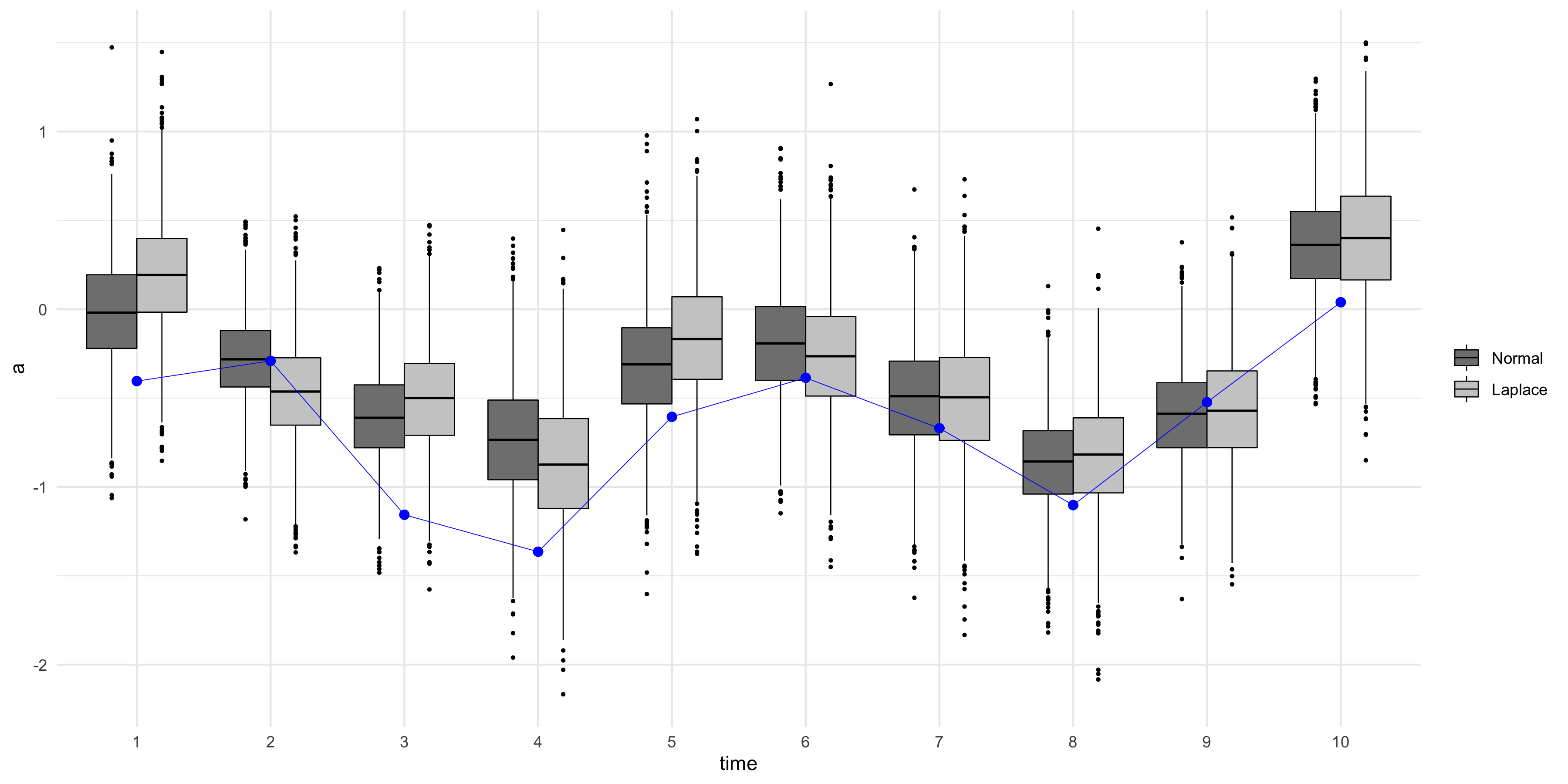}
		\includegraphics[width=0.62\textwidth]{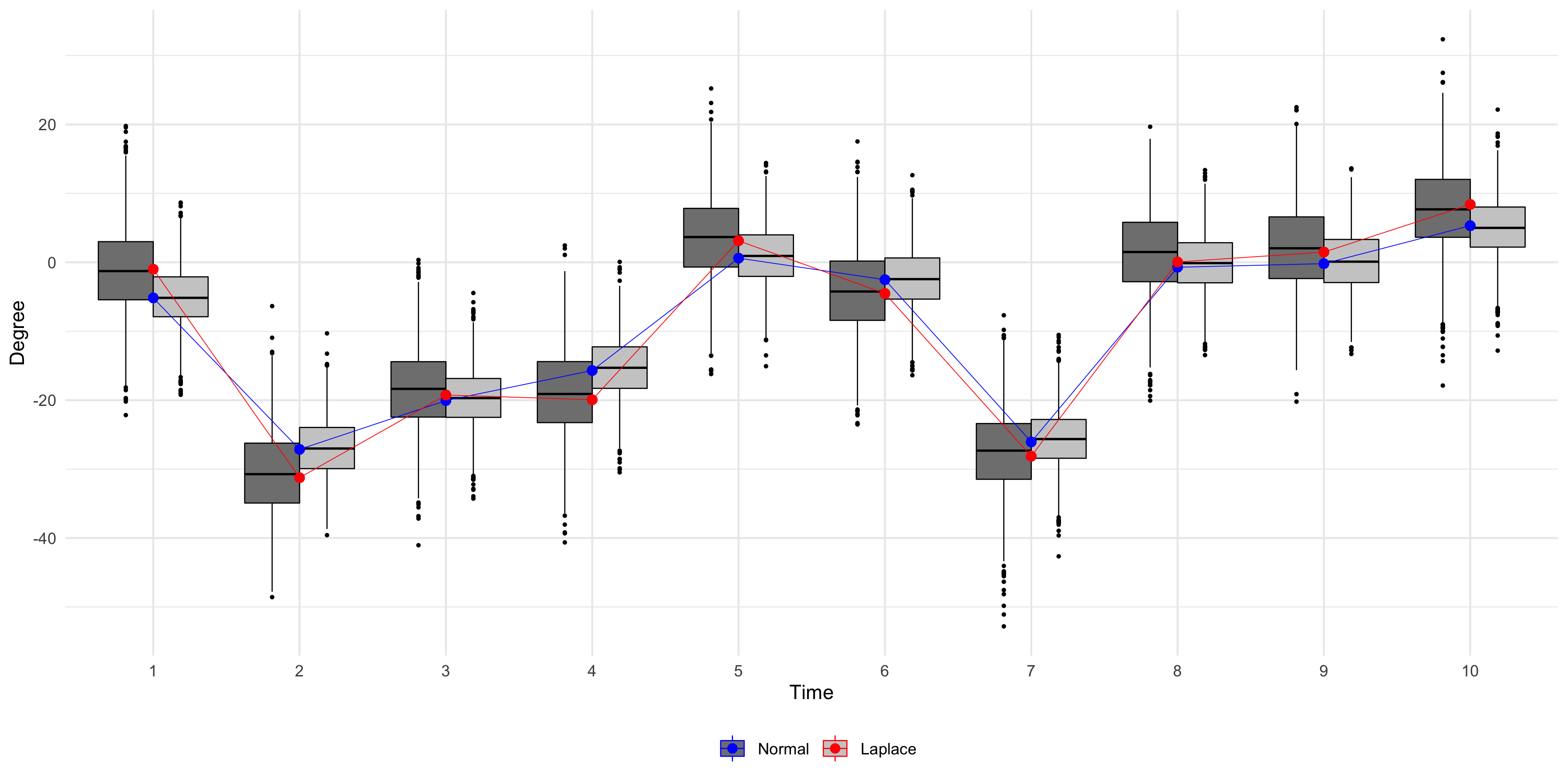}	
	\end{center}
		\caption{Boxplots of posterior samples of parameter estimates for the fixed effect (\textit{upper}) and the additive effect for node 1 (\textit{middle}), and posterior predictive degree statistics for node 1 (\textit{lower}) shown with the dots representing true value of parameters or observed statistics. The observed degree depends on the error distribution so the true values are slightly different in the lower panel.}
		\label{figure:PPC}
\end{figure}

\section{Posterior Summaries on GP Parameters}\label{appendix: Posterior summaries}
\begin{table}[H]
	\centering
	\begin{tabular}{ |c|ccccccccc|} 
		\hline
		$\kappa$ &  $\kappa_{intercept}^\beta$&	$\kappa_{log(distance)}^\beta$& $\kappa_{polity}^\beta$ &   $\kappa_{alliance} ^\beta$&  $\kappa_{Trade/GDP} ^\beta$ &  $\kappa_{language}^\beta$ & $\kappa^\theta$ & $\kappa_1^u$ &$\kappa_2^u$ \\
		\hline
		2.5\% & 27.883 & 1.876 & 53.448 & 1.432& 136.582 & 1.352 & 3.567 & 9.251 & 0.762\\
		25\% & 44.995 & 3.160 & 109.687 & 2.469 & 307.791 & 2.167 & 3.836 & 10.272 & 0.831 \\
		 50\% & 60.641 & 4.301 & 162.852 & 3.214 & 486.600 & 2.797 & 3.989 & 10.804 & 0.866\\
		 mean & 68.514 & 4.817 & 194.444 & 3.587 & 650.589 & 3.085 & 4.001 & 10.874 & 0.867\\
		75\% & 82.561 & 5.761 & 234.342 & 4.262 & 786.030 & 3.638 & 4.156 & 11.452 & 0.904 \\
		97.5\% & 154.887 & 11.173 & 521.416 & 7.848 & 2076.781 & 6.614& 4.482 & 12.693 & 0.971\\
		\hline
	\end{tabular}
	\caption {Posterior distribution of $\kappa$.}
	\label{table:kappas}
\end{table}
\begin{table}[H]
	\centering
	\begin{tabular}{ |c|ccccccccc|} 
		\hline
		$\tau$ &  $\tau_{intercept}^\beta$&	$\tau_{log(distance)}^\beta$ & $\tau_{polity}^\beta $&   $\tau_{alliance}^\beta$ &  $\tau_{Trade/GDP} ^\beta$ &  $\tau_{language}^\beta$ & $\tau^\theta$ & $\tau_1^u$ &$\tau_2^u$ \\
		\hline
		2.5\% & 403.820 & 0.161 & 0.087 & 0.978 & 0.091 & 0.420 & 29.211 & 6.870 & 4.061\\
		25\% &  605.178 & 0.226 & 0.146 & 1.397 & 0.145 & 0.610& 31.094 & 7.887 & 4.683\\
		 50\% & 780.451 & 0.281 & 0.194 & 1.720& 0.196 & 0.748 & 32.208 & 8.520 & 50.046\\
		 mean & 881.177 & 0.311 & 0.220 & 1.876 & 0.228 & 0.807 & 32.276 & 8.617 & 5.103\\
		75\% & 1039.767 & 0.360 & 0.264& 2.170 & 0.271 & 0.938 & 33.367 & 9.230 & 5.466 \\
		97.5\% & 1904.540 & 0.623 & 0.503 & 3.675 & 0.549 & 1.517 & 35.842 & 10.959 & 6.334\\
		\hline
	\end{tabular}
	\caption {Posterior distribution of $\tau$.}
	\label{table:taus}
\end{table}
\section{Posterior Predictive Checks}\label{appendix: PPC}
\begin{figure}[H]
	\begin{center}
	\includegraphics[width=0.515\textwidth]{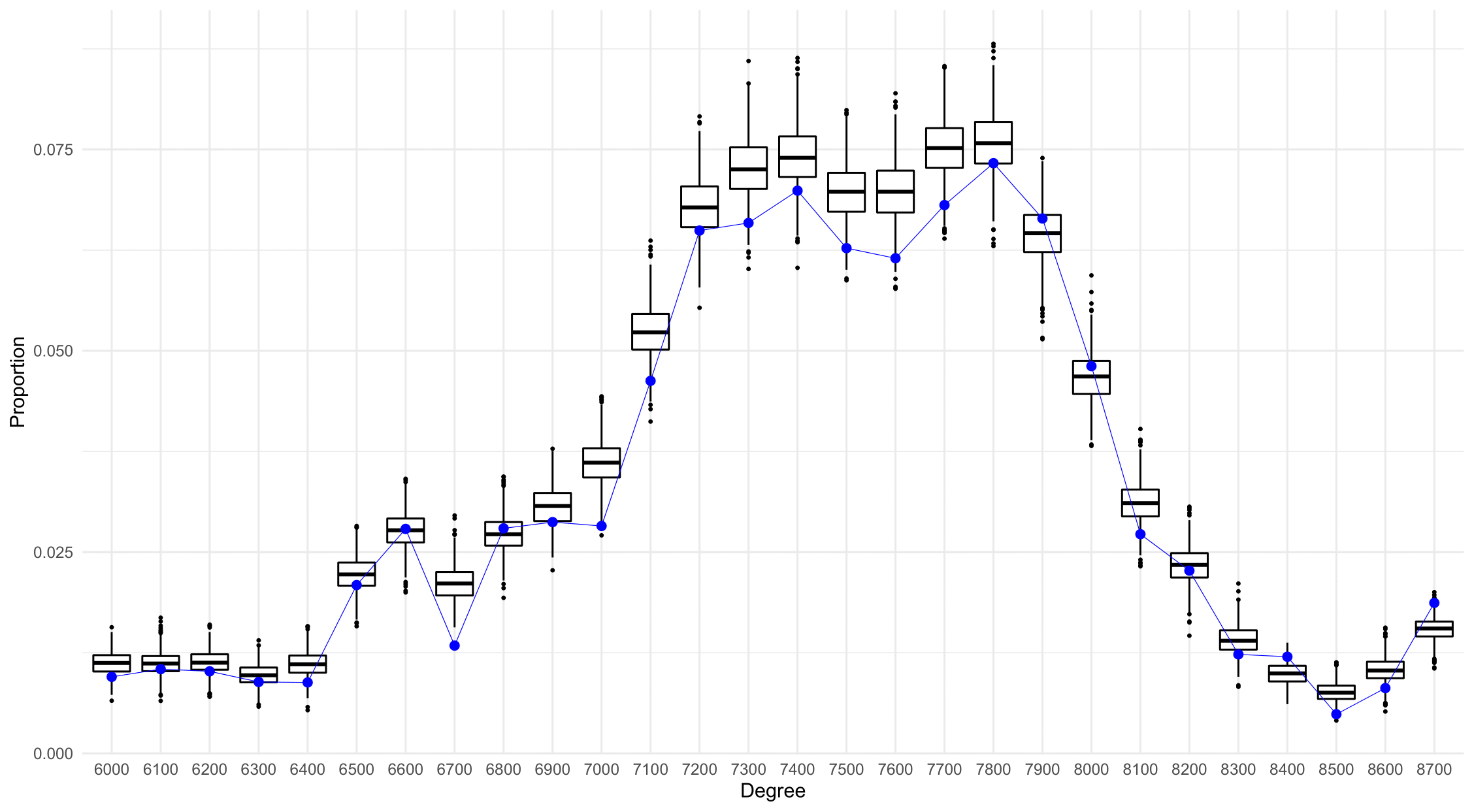}	
	\includegraphics[width=0.515\textwidth]{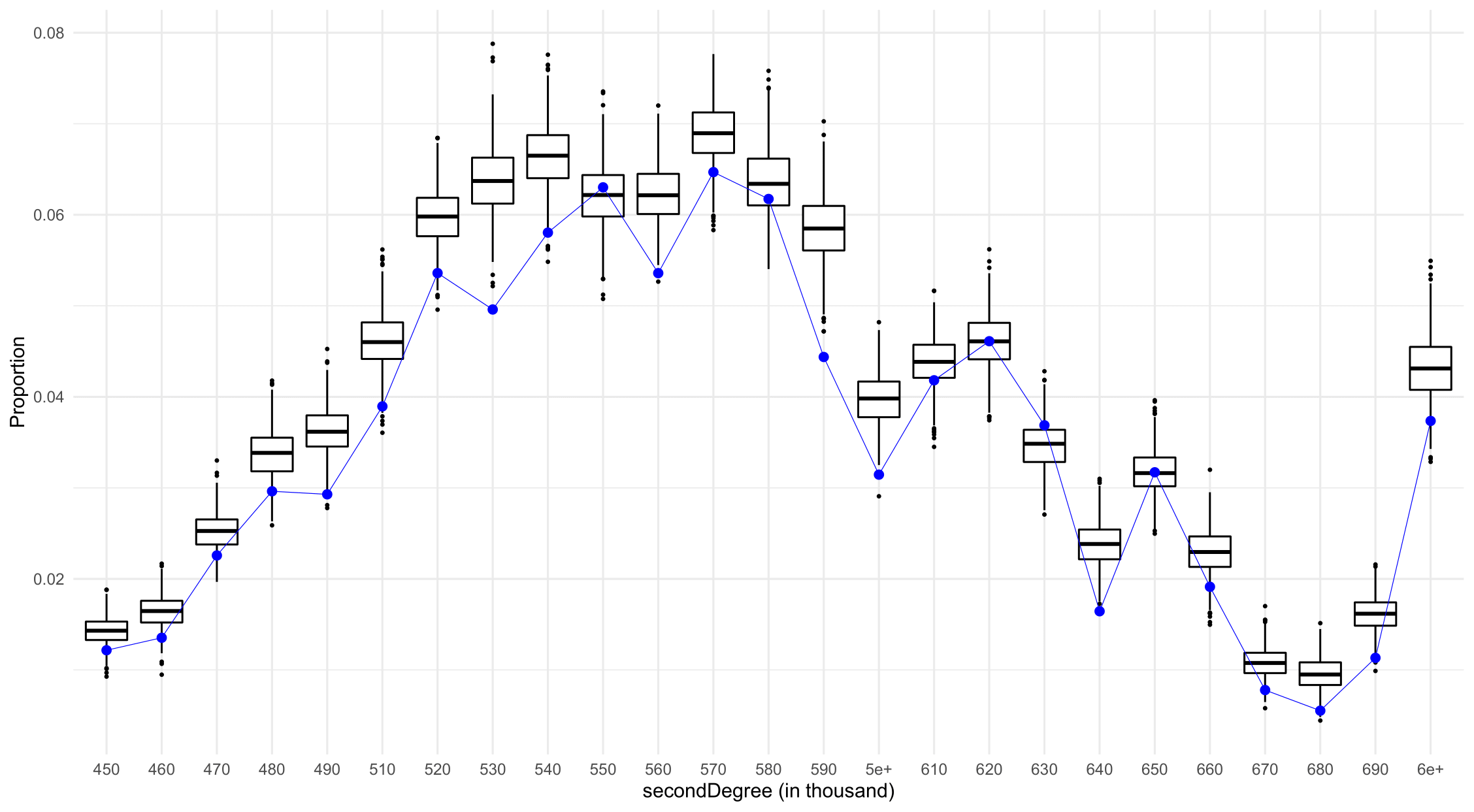}
	\includegraphics[width=0.515\textwidth]{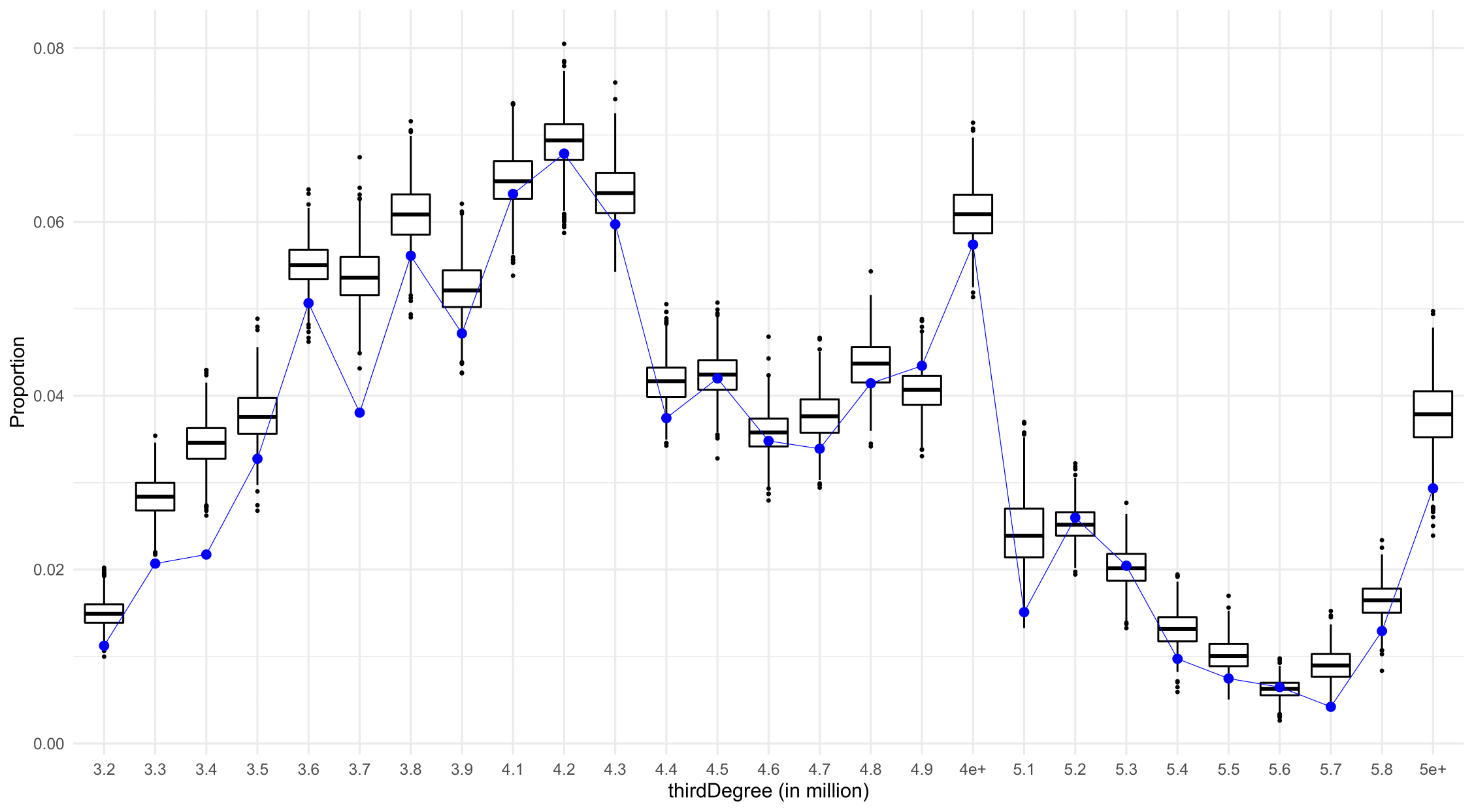}	
	\end{center}
		\caption {Posterior predictive plots of the overall degree distributions aggregating all nodes and timepoints: the first (\textit{upper}), second (\textit{middle}), and third moments (\textit{lower}) shown with the dots representing observed statistics.}
		\label{figure:PPC}
\end{figure}

\begin{figure}[H]
	\begin{center}
	\includegraphics[width=1\textwidth]{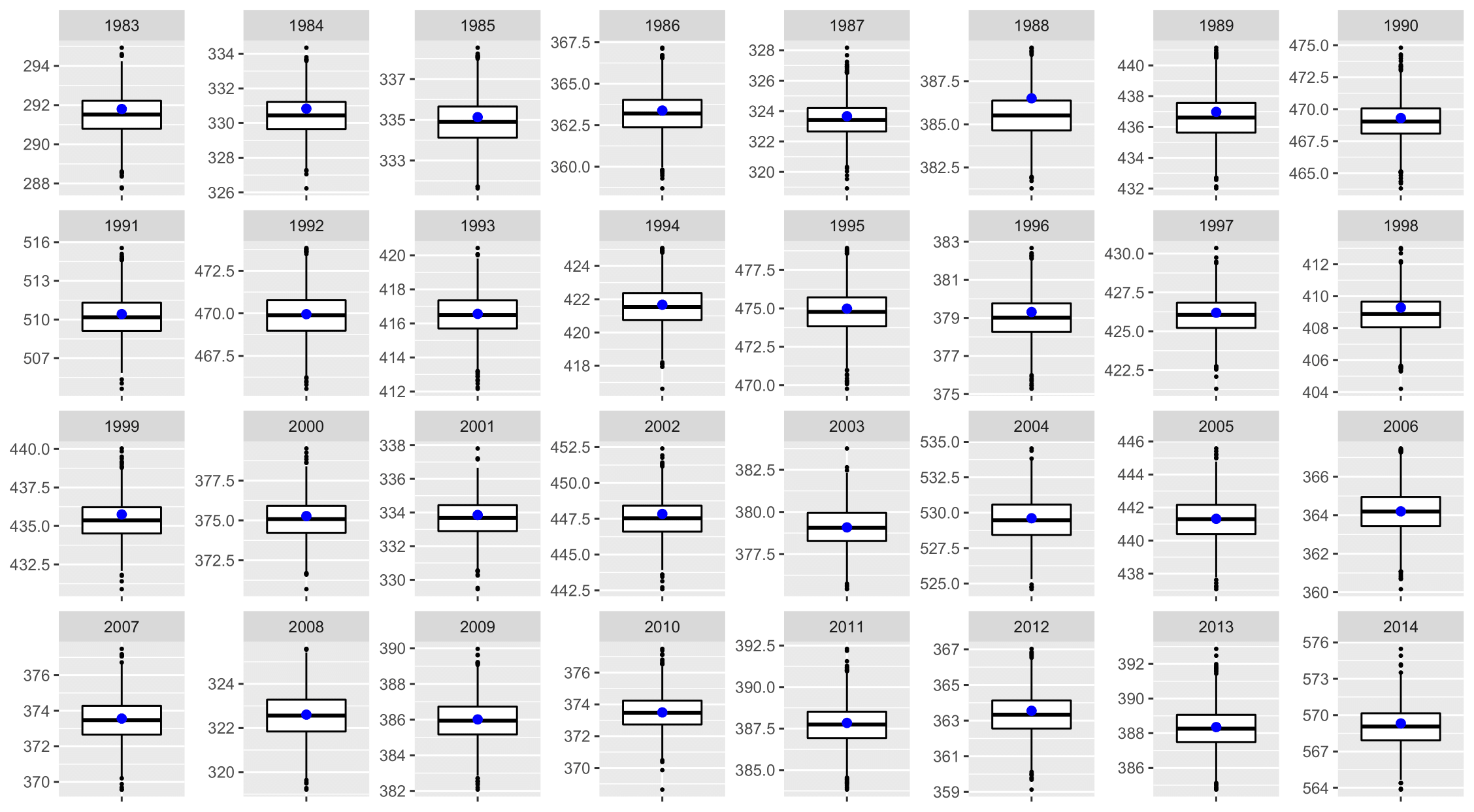}
	\end{center}
		\caption {Posterior predictive plots based on transitivity statistics (defined as the sum of diagonals from third moments aggregating all nodes), shown with the dots representing the observed statistics.}
		\label{figure:PPC2}
\end{figure}

\end{appendix}
\bibliographystyle{apalike}
\bibliography{BominBib2}